\pdfoutput=1
\documentclass[aps,pra,superscriptaddress,nofootinbib,10pt,a4paper]{revtex4-2}
\usepackage{amsmath,amssymb,amsfonts,graphicx,bm} 
\usepackage[colorlinks=true,allcolors=blue]{hyperref}
\usepackage[caption=false]{subfig}
\usepackage{tikz}
\usetikzlibrary{decorations.pathreplacing}

\usepackage{geometry}
\usepackage{float}
\setcitestyle{numbers,sort&compress}

\usepackage{subcaption}
\numberwithin{equation}{section}

\graphicspath{{ChargedChiralFigures/}}

\begin{document}

\title{Magnetic Q-balls}

\author{A.J. Balseyro Sebastian}
\affiliation{ Departamento de Matematica Aplicada, Universidad de Salamanca, SPAIN}
\author{Keisuke Ohashi}
\affiliation{Department of Physics $\&$ Research and Education Center for Natural Sciences, Keio University, 4-1-1 Hiyoshi, Yokohama, Kanagawa 223-8521, Japan}
\author{Muneto Nitta}
\affiliation{Department of Physics $\&$ Research and Education Center for Natural Sciences, Keio University, 4-1-1 Hiyoshi, Yokohama, Kanagawa 223-8521, Japan}
\address{International Institute for Sustainability with Knotted Chiral Meta Matter (WPI-SKCM$^2$), Hiroshima University, 1-3-1 Kagamiyama, Higashi-Hiroshima, Hiroshima 739-8531, Japan}
\date{September 25, 2026}

	\begin{abstract}
		We study  charged soliton branches in 
        quasi-one-dimensional
        chiral magnetic systems with
		 Dzyaloshinskii--Moriya (DM)  interaction, easy-axis anisotropy, and Zeeman coupling. The same static magnetic functional is equipped with two different dynamical completions: an antiferromagnetic model with second-order time derivatives and a ferromagnetic model with Berry-phase dynamics.
        On the helical branch selected by the static DM interaction, the problem reduces to an analytically tractable one-dimensional system for the polar angle of the order parameter.
		We derive the existence conditions for polar Q-balls from the curvature of the reduced effective potential and the presence of a nonzero turning point. In the antiferromagnetic case, the allowed frequency window is symmetric and can be completely closed by the combined effect of the DM coupling and the Zeeman field. At zero Zeeman field, the same reduction also supports antiferromagnetic Q-kinks, for which we obtain explicit profiles, charges, energies, and reduced-sector fission criteria. In the ferromagnetic case, the Berry phase makes the rotation frequency act as a shifted Zeeman field. As a result, north- and south-pole charged droplets are selected by opposite signs of the shifted rotation. We also show that a formal pole-to-pole solution of the ferromagnetic mechanical problem does not generally correspond to a finite-energy magnetic soliton, because the Berry term does not renormalize the physical Hamiltonian.
		These results clarify how charged-soliton mechanisms depend on the underlying magnetic
		dynamics, even when the static chiral energy is the same.
	\end{abstract}

    \maketitle

	\section{Introduction}

Non-topological solitons stabilized by a conserved Noether charge constitute one of the most remarkable mechanisms for localized field configurations.
The prototypical example is the Q-ball proposed by Coleman
\cite{Coleman:1985ki}; For reviews and textbook accounts, see
Refs.~\cite{Lee:1991ax,Nugaev:2019vru,Shnir:2018yzp}.
Related non-topological scalar solitons had been studied before
Coleman's formulation \cite{Friedberg:1976me}.
See also, e.~g.~ Refs.~\cite{
Kusenko:1997ad,Multamaki:1999an,
Bowcock:2008dn,Tsumagari:2008bv,
Alonso-Izquierdo:2023xni,Alonso-Izquierdo:2023hrr} 
for further studies.
Besides their intrinsic field-theoretical interest, Q-balls have been studied
in supersymmetric extensions of the Standard Model
\cite{Kusenko:1997zq,Enqvist:1997si}
and in cosmology
\cite{Frieman:1988ut,Enqvist:2003gh}.
They arise naturally in connection with the Affleck--Dine mechanism
\cite{Affleck:1984fy}, including the formation of Q-balls through the
fragmentation of the Affleck--Dine condensate
\cite{Kasuya:1999wu},
and have also been proposed as possible dark-matter candidates
\cite{Kusenko:1997si}.
More generally, internal rotation can endow topological solitons with
a Noether charge, giving rise to isospinning solitons such as Q-kinks \cite{Abraham:1992vb,Abraham:1992qv,Gauntlett:2000ib},
Q-lumps \cite{Leese:1991hr,Abraham:1991ki,Naganuma:2001pu,Amari:2024pnw}, 
isospinning Skyrmions 
\cite{Battye:2014qva,Battye:2005nx},
and 
isospinning Hopfions 
\cite{Harland:2013uk,Battye:2013xf}.
Isospinning composite solitons are also constructed in Refs.~\cite{Eto:2006pg,Eto:2005sw,Eto:2007uc,Nitta:2012wi}.  
In $1+1$ dimensional relativistic field theories, 
Q-balls are localized non-topological configurations supported by a conserved $U(1)$ Noether charge, 
Q-kinks interpolate between distinct vacua while simultaneously carrying a conserved Noether charge
\cite{Abraham:1992vb,Abraham:1992qv,Gauntlett:2000ib}.
They are commonly understood through an effective one-dimensional mechanical problem:
a uniform internal rotation modifies the effective potential, and localized charged solitons exist only when the resulting mechanical potential possesses appropriate turning points and asymptotic behavior.

An interesting question is whether the same charged-soliton mechanism can
also be realized in condensed-matter systems.
One example is provided by Q-balls in superfluid $^3$He
\cite{Bunkov:2007fe,Autti:2017eya}.
The main interest of the present work is in magnetic systems.
Localized magnetic textures are central objects in nonlinear spin systems.
Depending on the symmetry and energetics of the material, they appear as
domain walls, helices, vortices, droplets, and skyrmions. 
Magnetic systems provide a particularly interesting arena because the same
static magnetic energy functional may admit fundamentally different
dynamical completions depending on the underlying magnetic order.
Antiferromagnetic dynamics is naturally second order in time and is
therefore closely related to relativistic nonlinear sigma models
\cite{Haldane:1983ru,Fradkin:2013anc,KOSEVICH1990117},
whereas ferromagnetic dynamics is first order and governed by the Berry
phase.
Since charged solitons are intrinsically time-dependent objects, this
difference in dynamics can modify not only the properties of the solutions
themselves but also the very conditions for their existence.

Among magnetic systems, chiral magnets provide a particularly attractive setting.
The Dzyaloshinskii--Moriya (DM) interaction
\cite{Dzyaloshinsky:1958cyg,Moriya:1960zz}
stabilizes spatially twisted spin textures and selects a preferred chirality.
As a consequence, chiral magnets exhibit a rich variety of localized structures, including magnetic domain walls \cite{togawa2012chiral,KISHINE20151,PhysRevB.97.184303,PhysRevB.65.064433,Ross:2020orc}, magnetic skyrmions
\cite{1989JETP...68..101B,1995JETPL..62..247B,
Rossler:2006,
doi:10.1126/science.1166767,doi:10.1038/nature09124,doi:10.1038/nphys2045,Lin:2014ada,Han:2010by,Ross:2020hsw,Nagaosa2013},
monopoles \cite{tanigaki2015,fujishiro2019topological}, 
Hopfions \cite{Sutcliffe:2018vcb} 
and 
instantons \cite{Hongo:2019nfr}.  
See Ref.~\cite{GOBEL20211} 
as a review.
Recently, composite configurations involving two or more types of
topological solitons have also been investigated, including 
magnetic domain-wall skyrmions 
\cite{Nagase:2020imn,li2021magnetic,Yang:2021,PhysRevB.99.184412,Kim:2017lsi,Lee:2022rxi,PhysRevB.102.094402,Kuchkin:2020bkg,Ross:2022vsa,Amari:2023gqv,PhysRevB.109.014404,Amari:2023bmx,Gudnason:2024shv,Leask:2024dlo,Amari:2024jxx,Gudnason:2025gqs},
a domain wall network \cite{Lee:2024lge}, 
a D-brane soliton (skyrmion string ending on a N\'eel domain wall) \cite{Gudnason:2025inp} 
and so on.
In particular, 
magnetic skyrmions 
have attracted considerable attention as promising information carriers for magnetic data-storage technologies \cite{Nagaosa2013}.

Another important point of comparison is provided by localized
precessing magnetic solitons, 
which have a long history in both
ferromagnets and antiferromagnets \cite{KOSEVICH1990117}.
More recently, magnetic droplet solitons have been extensively studied
in ferromagnetic systems \cite{ahlberg2024magnetic,Jiang:2024}.
They include both conservative Landau--Lifshitz solitons and
driven-dissipative states sustained by spin torque.
These studies provide an important background for asking how the
charged-soliton mechanism is modified by the DM-induced helical twist
and by the qualitatively different dynamics of antiferromagnets and
ferromagnets.
It is therefore natural to ask whether chiral magnets can also support charged solitons analogous to relativistic Q-balls.

In this work, we answer this question by studying charged solitons in chiral magnets with DM interaction.
The order parameter is a unit vector field taking values on $S^2$, and the static energy consists of exchange, DM interaction, easy-axis anisotropy, and a Zeeman term.
We work on the helical branch selected by the static DM interaction, where the
spatial pitch is fixed to its preferred value. Adding a uniform internal
rotation on this branch reduces the problem to an analytically tractable
one-dimensional theory, 
which is 
the double sine-Gordon model \cite{PhysRevB.27.474}, see also Refs.~\cite{PhysRevB.65.064433,Ross:2020orc,Amari:2024jxx}. 
This framework allows us to compare two distinct dynamical completions of the same static magnetic energy, namely an antiferromagnet with second-order dynamics and a ferromagnet governed by the Berry phase.

We show that the charged-soliton mechanisms are fundamentally different in the two cases.
In the antiferromagnetic model, the DM coupling $D$ and the internal
rotation frequency $\omega$ enter through the combination
$D^2+\omega^2$, which controls the curvature and turning points of the
reduced potential and hence the existence of Q-balls and Q-kinks. 
In contrast, in the ferromagnetic model the Berry phase makes the rotation frequency act as a shift of the effective Zeeman field.
Consequently, the direction of rotation becomes physically relevant, and the north- and south-pole charged branches are selected by opposite signs of the shifted frequency.

The paper is organized as follows.
In Sec.~\ref{sec:model}, we formulate the chiral magnetic model, introduce the
antiferromagnetic and ferromagnetic dynamical completions, and discuss
the ground-state structure.
In Sec.~\ref{sec:static}, before turning to charged rotating solitons, we analyze the
static helical sector.
In Sec.~\ref{sec:stationary}, we formulate the stationary rotating helical sector common
to the subsequent analysis.
Sections~\ref{sec:antiferro-Q} and \ref{sec:ferro-Q} are devoted to antiferromagnetic and ferromagnetic
charged solitons, respectively.
In Sec.~\ref{sec:comparison}, we compare the two dynamical mechanisms and summarize their
existence regions.
Finally, Sec.~\ref{sec:summary} contains our discussion and outlook.
In Appendix \ref{app:bulk-model}, the rotating helical reduction of three dimensional chiral magnets 
is used to derive the
	effective one-dimensional problems and their conserved charges.
In Appendix \ref{app:pole-crossing}, we discuss extended pole-crossing solutions that pass through the opposite pole and return to the same physical vacuum.

    \section{The model and the ground states}
    \label{sec:model}

\subsection{Chiral magnetic model}
    \label{subsec:chiral-model}

Low-energy phenomena in chiral magnets can be described by a continuum
theory formulated in terms of a three-component unit vector
$\bm{n}=(n_x,n_y,n_z)$ taking values on $S^2$.
The physical interpretation of $\bm{n}$ depends on the underlying magnetic
order.  In a ferromagnet, $\bm{n}$ represents the direction of the local
magnetization,
\begin{equation}
    \bm{n}=\frac{\bm{M}}{M_s},
\end{equation}
where $\bm{M}$ is the magnetization and $M_s$ is the saturation
magnetization.  In an antiferromagnet, the corresponding low-energy
order parameter is the normalized N\'eel vector, which describes the
staggered magnetization of the two sublattices.  Although their dynamical
theories are qualitatively different, the two systems can be described
by the same static chiral energy functional in terms of the unit vector
$\bm{n}$.

The continuum model considered in this paper is governed by a static
energy containing exchange, a chiral DM
interaction, an easy-axis anisotropy, and a Zeeman term,
\begin{equation}
        {\cal E}_{\rm stat}=\frac{1}{2} \nabla{\mathbf n} \cdot \nabla {\mathbf n}
        - D \, {\mathbf n}
        \cdot 
        (\nabla \times {\mathbf n})
+\mathbf{H}\cdot {\mathbf n} + \frac{m^2}{2} \left(1-n_{z}^{2}\right).
\label{eq:bulk_theory}
\end{equation}
We mainly consider quasi-one-dimensional 
system along the $z$ direction 
so that the easy-axis potential is inline.
We take the effective field along the easy axis, $\bm H=(0,0,h)$.
Here the linear term denoted by $\bm H$ should be understood more generally
as an effective field conjugate to $n_z$; in the ferromagnetic realization
it is the usual Zeeman field.  This formulation allows us to isolate the
role of the dynamics by comparing antiferromagnetic and ferromagnetic
dynamical completions of the same static theory. 
We formulate the model directly in the \(1+1\)-dimensional Minkowski space, with
coordinates \((t,z)\) and metric
	$ds^2=dt^2-dz^2$.
The order parameter  \(\mathbf n(t,z)\in S^2\) 
can be represented in the spherical coordinates as
\begin{equation}
	\mathbf n
	=
	\left(
	\sin\theta\cos\varphi,\,
	\sin\theta\sin\varphi,\,
	\cos\theta
	\right),
\end{equation}
where \(\theta=\theta(t,z)\) and \(\varphi=\varphi(t,z)\). The target-space metric is
$ds_{S^2}^2=d\theta^2+\sin^2\theta\,d\varphi^2$.
 In the conventions used in this paper, the static energy 
 in Eq.~(\ref{eq:bulk_theory})
 reduces to 
\begin{equation}
	\mathcal E_{\rm stat}
	=
	\frac{1}{2}\theta_z^2
	+
	\frac{1}{2}\sin^2\theta\,\varphi_z^2
	+
	D\sin^2\theta\,\varphi_z
	+
	\frac{m^2}{2}\sin^2\theta
	+
	h\cos\theta .
\end{equation}
Equivalently, completing the square in the chiral derivative,
\begin{equation}
\mathcal E_{\rm stat}
=
\frac{1}{2}\theta_z^2
+
\frac{1}{2}\sin^2\theta\,(\varphi_z+D)^2
+
\frac{\delta}{2}\sin^2\theta
+
h\cos\theta ,
\qquad
\delta\equiv m^2-D^2 .
\label{eq:delta}
\end{equation}
This form makes explicit that the DM interaction selects a preferred helical pitch, while also lowering the effective easy-axis coefficient from \(m^2\) to \(\delta=m^2-D^2\).

The potential part will therefore be written as
\begin{equation}
	V(\theta)=\frac{m^2}{2}\sin^2\theta+h\cos\theta .
\end{equation}
For \(h=0\), the two polar points \(\theta=0,\pi\) are degenerate vacua. For \(h\neq0\), one of the two poles is energetically preferred, and this will be important in distinguishing Q-ball and Q-kink sectors.

We consider two dynamical completions of the same static chiral magnetic functional. In the antiferromagnetic case the time evolution is second order, as in a relativistic nonlinear sigma model on \(S^2\).
The Lagrangian is
\begin{equation}
	\mathcal L_{\rm AF}
	=
	\frac{1}{2}
	\left(
	\theta_t^2+\sin^2\theta\,\varphi_t^2
	\right)
	-
	\mathcal E_{\rm stat}.
\end{equation}
Equivalently,
\begin{align}
	\mathcal L_{\rm AF}
	=&\;
	\frac{1}{2}
	\left(
	\theta_t^2+\sin^2\theta\,\varphi_t^2
	\right)
	-
	\frac{1}{2}
	\left(
	\theta_z^2+\sin^2\theta\,\varphi_z^2
	\right)-D\sin^2\theta\,\varphi_z
	-
	V(\theta).
\end{align}
The corresponding Hamiltonian density is
\begin{equation}
	\mathcal H_{\rm AF}
	=
	\frac{1}{2}
	\left(
	\theta_t^2+\sin^2\theta\,\varphi_t^2
	\right)
	+
	\mathcal E_{\rm stat}.
\end{equation}

In the ferromagnetic case the quadratic kinetic term is replaced by a Berry-phase term. A Berry gauge regular at the north pole is
\begin{equation}
	\mathcal L_B^{(N)}
	=
	\mu(\cos\theta-1)\varphi_t .
    \label{eq:Berry-north}
\end{equation}
The ferromagnetic Lagrangian density in this gauge is
\begin{equation}
	\mathcal L_{\rm FM}^{(N)}
	=
	\mu(\cos\theta-1)\varphi_t
	-
	\mathcal E_{\rm stat}.
     \label{eq:gauge-N}
\end{equation}
Equivalently,
\begin{align}
	\mathcal L_{\rm FM}^{(N)}
	=&\;
	\mu(\cos\theta-1)\varphi_t
	-
	\frac{1}{2}
	\left(
	\theta_z^2+\sin^2\theta\,\varphi_z^2
	\right)
	-D\sin^2\theta\,\varphi_z
	-
	V(\theta).
\end{align}
A Berry gauge regular at the south pole is instead
\begin{equation}
	\mathcal L_B^{(S)}
	=
	\mu(\cos\theta+1)\varphi_t .
    \label{eq:gauge-S}
\end{equation}
The two gauges differ by a total derivative,
\begin{equation}
	\mathcal L_B^{(S)}-\mathcal L_B^{(N)}
	=
	2\mu\varphi_t .\label{eq:gauges}
\end{equation}
They lead to the same local equations of motion, since they differ only by a
total time derivative. The choice of Berry gauge will become relevant below
when the conserved charge of a rotating configuration is defined relative to a
chosen asymptotic pole.
The ferromagnetic Hamiltonian density is independent of the Berry term:
\begin{equation}
	\mathcal H_{\rm FM}
	=
	\mathcal E_{\rm stat}.
\end{equation}
Thus the Berry phase fixes the symplectic structure and the conserved charge, but it does not contribute directly to the physical Hamiltonian.

The corresponding one-dimensional antiferromagnetic and ferromagnetic actions are
\begin{equation}
	S_{\rm AF}
	=
	\int dt\,dz\;\mathcal L_{\rm AF},
	\qquad
	S_{\rm FM}
	=
	\int dt\,dz\;\mathcal L_{\rm FM}.
\end{equation}

The effective model above can be obtained from the three-dimensional chiral magnetic functional by restricting to longitudinal configurations depending only on \(t\) and \(z\). The full \(3+1\)-dimensional formulation in cylindrical coordinates, including the bulk DM density, is given in Appendix~\ref{app:bulk-model}. The main text will use the one-dimensional formulation because the charged helical Q-ball and Q-kink sectors are defined inside this reduced model.

Both dynamical completions are invariant under global rotations around the
easy axis,
\begin{equation}
    \varphi\mapsto \varphi+\epsilon ,
\end{equation}
and the corresponding conserved quantities play the role of charges for the
stationary solitons studied below. In the antiferromagnetic model, the Noether current is
\begin{equation}
    J^t_{\rm AF}
    =
    \sin^2\theta\,\varphi_t,
    \qquad
    J^z_{\rm AF}
    =
    -\sin^2\theta\,(\varphi_z+D).
    \label{eq:AF-current-general}
\end{equation}
Thus the antiferromagnetic charge is
\begin{equation}
    Q_{\rm AF}
    =
    \int_{-\infty}^{\infty}
    \sin^2\theta\,\varphi_t\,dz .
    \label{eq:AF-charge-general}
\end{equation}

In the ferromagnetic model, the time component of the current is fixed by the Berry one-form, while the spatial component is again determined by the
gradient part of the magnetic energy. In the Berry gauge regular at the north pole,
\begin{equation}
    J^t_{{\rm FM},N}
    =
    \mu(\cos\theta-1),
    \qquad
    J^z_{\rm FM}
    =
    -\sin^2\theta\,(\varphi_z+D),
    \label{eq:FM-current-general}
\end{equation}
whereas in the Berry gauge regular at the south pole,
\begin{equation}
    J^t_{{\rm FM},S}
    =
    \mu(\cos\theta+1),
    \qquad
    J^z_{\rm FM}
    =
    -\sin^2\theta\,(\varphi_z+D).
\end{equation}
For localized fields approaching a specified pole, it is convenient to use the corresponding positive, vacuum-subtracted Berry charges
\begin{equation}
    Q^{\rm FM}_{N}
    =
    \mu
    \int_{-\infty}^{\infty}
    \left(1-\cos\theta\right)\,dz ,
    \qquad
    Q^{\rm FM}_{S}
    =
    \mu
    \int_{-\infty}^{\infty}
    \left(1+\cos\theta\right)\,dz .
    \label{eq:FM-polar-charges-general}
\end{equation}
These are the solid-angle charges measured in the Berry gauges adapted to the north and south poles. This already shows an important difference between the two dynamical completions. In the antiferromagnetic model the charge is proportional to \(\varphi_t\), so every static configuration has \(Q_{\rm AF}=0\). In the ferromagnetic model the Berry charge is instead geometric: a nontrivial static profile localized with respect to a polar vacuum carries a nonzero vacuum-subtracted solid-angle charge. In a fixed polar sector, the only configuration with zero Berry charge is the corresponding vacuum.

	\subsection{Static ground states: homogeneous versus helical phases}
	
	Before studying charged solitons, it is useful to identify the static ground states of the 
    quasi-one-dimensional chiral magnet. The antiferromagnetic and ferromagnetic theories introduced above are two dynamical completions of the same static functional. Therefore the homogeneous-versus-helical ground-state question is common to both models. Let us consider static configurations with constant polar angle and
longitudinal phase dependence,
\begin{equation}
    \theta=\text{constant},
    \qquad
    \varphi=\varphi(z) .
\end{equation}
The corresponding static energy density is
\begin{align}
    \mathcal E_{\rm stat}
    &=
    \frac{1}{2}\sin^2\theta(\varphi_z+D)^2
    +
    \frac{\delta}{2}\sin^2\theta
    +
    h\cos\theta ,
\end{align}
with \(\delta\) defined in Eq.~\eqref{eq:delta}. The pitch-dependent part is
minimized by the DM-preferred pitch
\begin{equation}
    \varphi_z=-D .
\end{equation}
 Using Eqs.~\eqref{eq:AF-current-general} and \eqref{eq:FM-current-general},
this is equivalently the zero-longitudinal-current branch,
\begin{equation}
    J^z=0 .
\end{equation}
Substituting this value gives
	\begin{equation}
		\mathcal E_{\rm stat}^{\rm hel}
		=
		\frac{\delta}{2}\sin^2\theta
		+
		h\cos\theta \,.
	\end{equation}
	
	This expression shows the basic competition. The anisotropy $m^2$ favors the polar states $\theta=0,\pi$, while the DM interaction lowers the energy of configurations with $\sin\theta\neq0$ by selecting a helical pitch. Thus $D$ weakens the effective easy-axis anisotropy. Depending on the sign of $\delta$:
    \begin{itemize}
		\item If $\delta>0$ the polar states are locally favored by the anisotropy. The homogeneous
		energies are
		\begin{equation}
			\mathcal E(0)=h,
			\qquad
			\mathcal E(\pi)=-h .
		\end{equation}
		For $\delta>h>0$ the south pole $\theta=\pi$ is the lower homogeneous vacuum, while for $-\delta< h<0$ the north pole $\theta=0$ is lower. For $h=0$ the two polar vacua are degenerate. 
        Indeed, the nature of the higher-energy polar state also depends on the local curvature of the potential. Expanding around the two poles, one finds that the higher-energy pole is a linearly stable false vacuum when
    \begin{align}
    |h|<\delta ,
    \end{align}
    whereas it becomes linearly unstable when
    \begin{align}
    |h|>\delta .
    \end{align}
    At the threshold \(|h|=\delta\), the quadratic curvature at one pole vanishes. The polar extremum is then marginal, and the leading local behavior is controlled by higher-order terms.
    
		\item If $\delta<0$ the DM interaction overwhelms the easy-axis anisotropy. The energy is lowered by moving away from the poles, and the ground state becomes a	helix. Indeed, the stationary condition gives
		\begin{equation}
			\cos\theta_0=\frac{h}{\delta}.
		\end{equation}
		Since $\delta<0$, this value $\theta_0$ corresponds to a minimum outside the poles only when $|h|<|\delta|$. In that regime the ground state is
		\begin{equation}
			\cos\theta_0=\frac{h}{\delta},
			\qquad
			\varphi=-Dz+\gamma \,,
		\end{equation}
		with a constant phase $\gamma\in \mathbb{R}$. The Zeeman field tilts the cone: increasing $|h|$ pushes the spin texture toward one of the poles. When $|h|\geq |\delta|$ the helix is saturated into a homogeneous polar state.
	\end{itemize}

	Thus the static phase structure can be summarized as follows:
	\begin{itemize}
		\item For weak enough
		DM coupling, $D^2<m^2$, the effective anisotropy is positive and the homogeneous polar states are the ground states.
		\item For strong enough DM coupling, $D^2>m^2$,	the effective anisotropy becomes negative and the system favors a helical helical ground state, unless the Zeeman field is large enough to force the order parameter back to a pole.
	\end{itemize}

	This ground-state discussion is independent of whether the subsequent dynamics is antiferromagnetic or ferromagnetic. The distinction between AF and FM appears only when time dependence is introduced. In the following sections we study charged excitations in the regime
\begin{equation}
	\delta=m^2-D^2>0,
\end{equation}
where the static ground states are the homogeneous polar vacua rather than helical states.

	\section{Static helical sector}\label{sec:static}

We now study static profiles on the zero-longitudinal-current helical branch
identified in the ground-state analysis. Since the antiferromagnetic and
ferromagnetic theories have the same static energy, this sector is common to
both dynamical completions.  Nevertheless, the charge interpretation of the same static profile is different in the two theories. In the antiferromagnetic model, static configurations are chargeless because the Noether charge in Eq.~\eqref{eq:AF-charge-general} is proportional to the time derivative of the internal phase. In the ferromagnetic model, by contrast, a non-homogeneous profile localized with respect to a fixed polar vacuum carries the vacuum-subtracted Berry charge defined in Eq.~\eqref{eq:FM-polar-charges-general}.

In the static one-dimensional
sector, the equation for the cyclic variable \(\varphi\) implies that \(J^z\)
is constant. At fixed nonzero \(J^z\), eliminating \(\varphi_z\) produces the
centrifugal contribution
\[
    \frac{(J^z)^2}{2\sin^2\theta},
\]
which is singular at the polar vacua. Therefore the finite-energy polar
profiles considered here lie on the zero-longitudinal-current branch selected
in the ground-state analysis. We take
\begin{equation}
    \theta=\theta(z),
    \qquad
    \varphi=-Dz+\gamma ,
\end{equation}
where \(\gamma\in\mathbb R\) is a constant phase. The remaining static problem
is a one-dimensional mechanical problem for the polar angle \(\theta(z)\).
 
With the preferred pitch imposed, the reduced static energy is
\begin{equation}
	E_{\rm stat}^{\rm hel}
	=
	\int dz
	\left[
	\frac{1}{2}\theta_z^2
	+
	V_{\rm eff}(\theta)
	\right],
\end{equation}
where
\begin{equation}
	V_{\rm eff}(\theta)
	=
	\frac{\delta}{2}\sin^2\theta
	+
	h\cos\theta
	+
	C \,.
\end{equation}
The constant \(C\in\mathbb R\) is chosen according to the asymptotic vacuum so that the energy density vanishes at spatial infinity. Equivalently, the static Euler--Lagrange equation can be written as the Newton equation
\begin{equation}\label{eq:SecondOrderEffective}
	\frac{d^2\theta}{dz^2}
	=
	\frac{dV_{\rm eff}}{d\theta},
\end{equation}
where \(z\) plays the role of mechanical time and \(-V_{\rm eff}\) is the mechanical potential. The additive constant \(C\) does not affect this second-order equation, but it fixes the zero of the first integral. For a finite-energy profile approaching an asymptotic pole, \(C\) is chosen so that \(V_{\rm eff}\) vanishes at that pole. This leads to the first-order equation
\begin{align}\label{eq:FirstOrderEffective}
\frac{1}{2}\theta_z^2=V_{\rm eff}(\theta).
\end{align}

\begin{figure}[h]
    \centering
    \includegraphics[width=\textwidth]{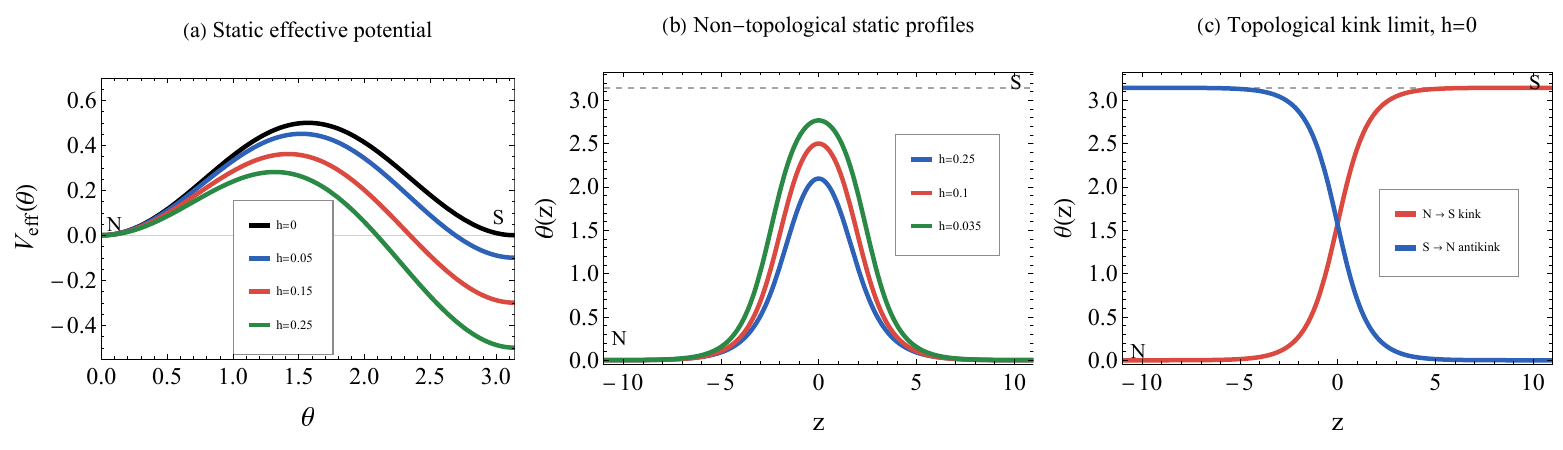}
    \caption{
    Static helical sector for \(\delta=1\). (a) Vacuum-subtracted effective potential around the north pole for different values of \(h\). 
    (b) Finite-\(h\) non-topological static profiles. As \(h\) decreases, the excursion approaches the opposite pole and the profile broadens.
    (c) At \(h=0\), the two polar vacua are degenerate and the limiting solutions are topological kink and antikink profiles.}
    \label{fig:static-helical-sector}
\end{figure}

\begin{figure}[h]
    \centering
    \begin{minipage}[t]{0.36\textwidth}
        \centering
        \includegraphics[width=\textwidth]{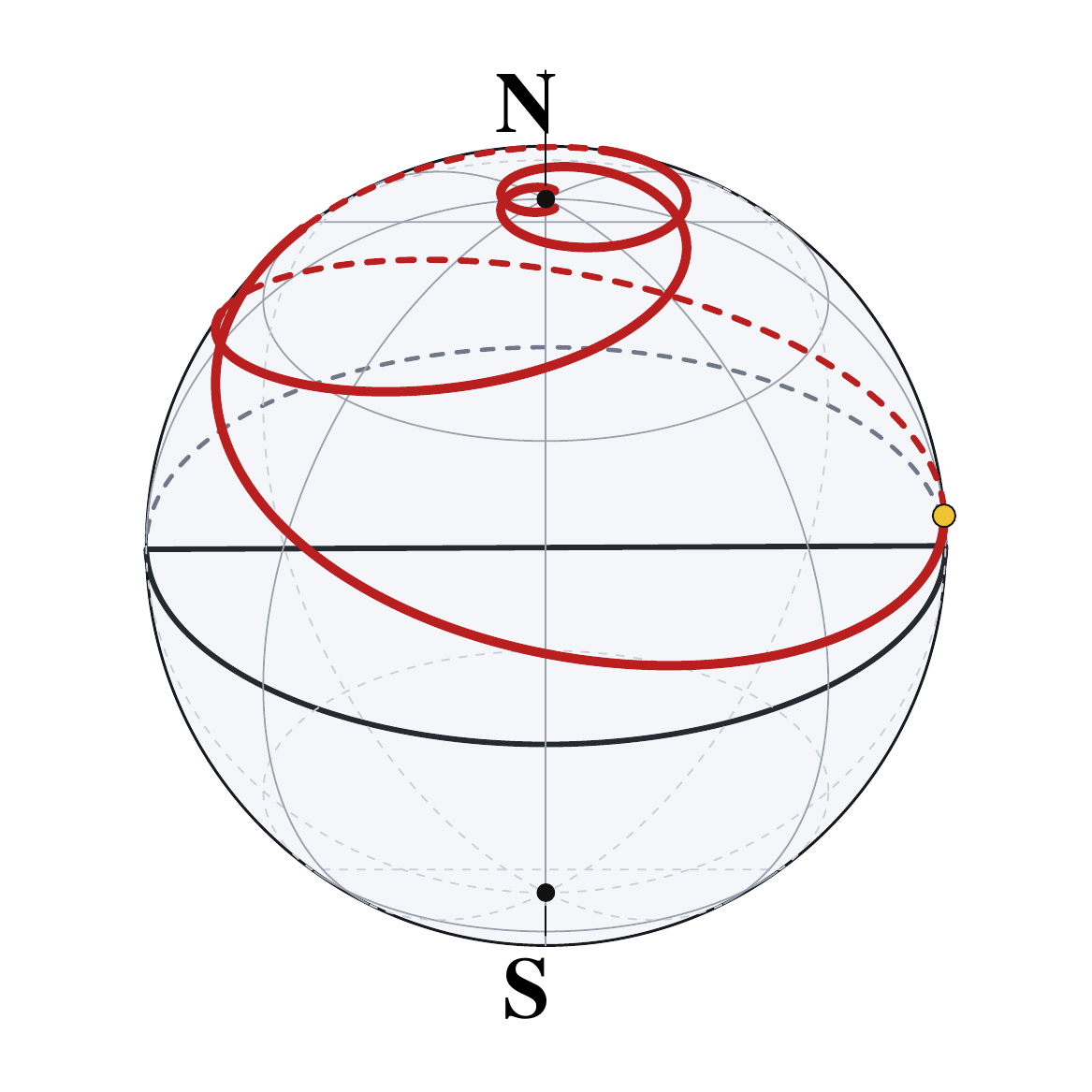}
        (a) Non-topological static profile.
        \label{fig:static-sphere-nontopological}
    \end{minipage}
    \hspace{0.05\textwidth}
    \begin{minipage}[t]{0.36\textwidth}
        \centering
    \includegraphics[width=\textwidth]{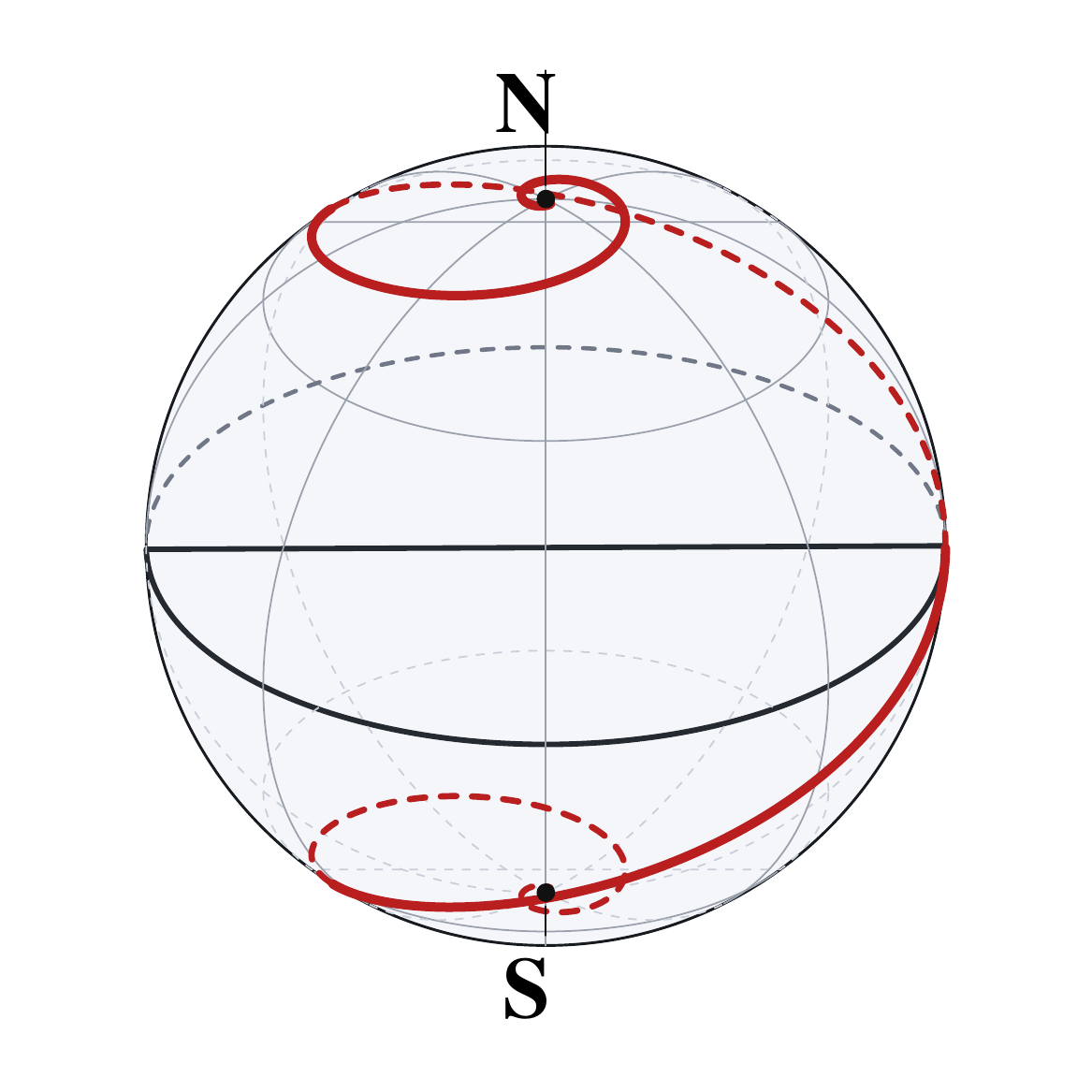}
        (b) Topological static kink.
        \label{fig:static-sphere-topological}
    \end{minipage}
    \caption{
    Static solutions represented as paths on the target sphere. Left: the non-topological profile leaves the north pole, reaches a turning point before the equator, and returns to the same vacuum along the same path. Right: at \(h=0\), the two polar vacua are degenerate and the topological kink connects the north and south poles. The antikink is the same curve with the opposite orientation.}
    \label{fig:static-sphere-solutions}
\end{figure}

When \(h=0\), the two poles remain degenerate vacua, as shown in
Fig.~\ref{fig:static-helical-sector} (a). Choosing \(C=0\), the static equation admits topological kink and antikink profiles connecting the two poles,
\begin{equation}
	\theta(z)
	=
	\arccos\left[
	\pm\tanh\left(\sqrt{\delta}\,(z-z_0)\right)
	\right],
	\qquad
	E=2\sqrt{\delta},
\end{equation}
where \(z_0\in\mathbb R\) is the center of the kink, see Fig.~\ref{fig:static-helical-sector}\,(c) and 
Fig.~\ref{fig:static-sphere-solutions}\, (b). 
This configuration was studied in Ref.~\cite{Borisov1985}.

For \(h\neq0\), the two poles are no longer degenerate. Localized non-topological static profiles can still be obtained around a polar local minimum when the corresponding curvature is positive and a turning point exists. For example, around the north pole one subtracts the north-pole energy by choosing \(C=-h\), giving
\begin{equation}
	V_{\rm eff}^{(N)}(\theta)
	=
	\frac{\delta}{2}\sin^2\theta
	+
	h(\cos\theta-1).
\end{equation}
For \(0<h<\delta\), the north pole is a local minimum of the static potential although it is not the global vacuum. The corresponding non-topological profile is
\begin{equation}
	\tan\frac{\theta_N(z)}{2}
	=
	\sqrt{
		\frac{\delta-h}{h}
	}\,
	\operatorname{sech}
	\left[
	\sqrt{\delta-h}\,(z-z_0)
	\right],
	\qquad
	E
	=
	4\sqrt{\delta-h}
	-
	\frac{4h}{\sqrt{\delta}}\,
	\operatorname{arctanh}
	\sqrt{\frac{\delta-h}{\delta}} ,
\end{equation}
see Fig.~\ref{fig:static-helical-sector}\,(b) and 
Fig.~\ref{fig:static-sphere-solutions}\,(a).
In the limit \(h\to0^+\), the turning point approaches the south pole. The non-topological profile therefore approaches a widely separated kink--antikink configuration, and its energy tends to twice the energy of the topological kink.

Similarly, a south-pole non-topological profile is obtained by subtracting the south-pole energy, \(C=h\). It exists for \(-\delta<h<0\), where the south pole is a local minimum but not the global vacuum. Thus the two non-topological branches are exchanged under
\[
	h\mapsto -h,
	\qquad
	\theta\mapsto \pi-\theta .
\]

The existence of these static profiles is controlled by the curvature of \(V_{\rm eff}\) at the poles,
\begin{equation}
	V''_{\rm eff}(0)=\delta-h,
	\qquad
	V''_{\rm eff}(\pi)=\delta+h .
\end{equation}
Therefore the following regimes appear:
\begin{itemize}
	\item If \(h=0\), the two polar vacua are degenerate and topological kinks interpolate between them.
	\item If \(0<|h|<\delta\), both poles are local minima, but only one is the global vacuum. The higher-energy pole supports a non-topological static profile with a finite turning point.
	\item If \(|h|>\delta\), one of the polar local minima is lost, and the corresponding non-topological branch disappears.
    \end{itemize}

There is also a second class of static finite-energy solutions if the meridional coordinate is continued beyond the fundamental interval \(0\leq\theta\leq\pi\) as discussed in the Appendix~\ref{app:pole-crossing}.
These extended pole-crossing profiles start and end at the same physical polar vacuum, but pass through the opposite pole along the meridian. For example, with north-pole subtraction,
\begin{align}
V_{\rm eff}^{(N)}(\theta)
=
2\sin^2\frac{\theta}{2}
\left[
\delta\cos^2\frac{\theta}{2}-h
\right].
\end{align}
At the opposite pole, \(\theta=\pi\), one has
\begin{align}
  V_{\rm eff}^{(N)}(\pi) = -2h .
\end{align}
Thus, for \(h<0\), the opposite pole is not a turning point or a forbidden point of the first-order flow, but a regular crossing point with nonzero \(\theta_z\). This allows an extended pole-crossing profile to pass through \(\theta=\pi\) and reach the equivalent point \(\theta=2\pi\). In the polar regime \(\delta>0\), the north-pole extended profile with \(\theta(-\infty)=0\) and \(\theta(+\infty)=2\pi\) therefore exists for \(h<0\), whereas the ordinary north-pole same-vacuum profile discussed above exists for \(0<h<\delta\). The south-pole extended profile is obtained by the reflection \(h\mapsto -h\), and therefore exists for \(h>0\). In this sense the extended profiles occur in the complementary Zeeman-field sector to the ordinary localized same-pole profiles.

These extended meridional solutions are finite-energy solutions of the same reduced static equation. While these extended meridional configurations initially appear to be kink-like solutions, the starting point \(\theta=0\) and the endpoint \(\theta=2\pi\) represent the same physical vacuum on \(S^2\). Consequently, the solution describes a closed loop in the target space. Since \(\pi_1(S^2)\simeq0\), such a loop has no topological protection and can be continuously deformed into the constant vacuum. This provides a geometric indication that the extended branch should not be treated as a protected kink sector.

The charge distinction mentioned at the beginning of this section is especially important for pole-to-pole profiles. In the antiferromagnetic theory, all static profiles in this section remain neutral because \(\varphi_t=0\). In the ferromagnetic theory, a localized same-pole profile has a finite vacuum-subtracted Berry charge in the polar gauge adapted to its asymptotic vacuum. A pole-to-pole static wall is different: its two spatial infinities approach different poles, so no single polar Berry gauge subtracts the vacuum charge density at both ends. Consequently, such a profile does not define an isolated finite localized ferromagnetic charge in one fixed polar gauge.

	\section{Stationary rotating helical sector}\label{sec:stationary}

We now add a uniform internal rotation while remaining on the same
zero-longitudinal-current helical branch. Thus
\begin{equation}
    \varphi(z,t)=-Dz+\omega t+\gamma,
    \qquad
    \theta=\theta(z),
    \label{eq:rotating-helical}
\end{equation}
where \(\omega\in\mathbb R\) is the angular velocity and \(\gamma\in\mathbb R\)
is a constant phase. This ansatz describes charged excitations of the
fixed-pitch helical branch.

Equations reduce then to a one-dimensional mechanical problem for \(\theta(z)\) of the form \eqref{eq:SecondOrderEffective}, with \(z\) playing the role of mechanical time. The fixed-frequency effective potentials are different in the antiferromagnetic and ferromagnetic models:
\begin{align}
	V_{\rm eff,AF}
	&=
	\frac{\alpha}{2}\sin^2\theta
	+
	h\cos\theta
	+
	C_1,
    \label{eq:EffectivePotentials-AF}
	\\
	V_{\rm eff,FM}
	&=
	\frac{\delta}{2}\sin^2\theta
	+
	\Lambda \cos\theta
	+
	C_2, 
    \label{eq:EffectivePotentials-FM}
\end{align}
with 
\begin{align}
 &   \alpha
		\equiv
		\delta-\omega^2
		=
		m^2-D^2-\omega^2 
        ,
        \label{eq:alpha}
        \\
&   
   \Lambda \equiv h-\mu\omega
   \label{eq:Lambda}
\end{align}
where \(C_1,C_2\in\mathbb R\) are chosen according to the asymptotic vacuum. Once again, the constants \(C_1\) and \(C_2\) do not affect the second-order equation, but they fix the zero of the first integral. Along a finite-energy mechanical orbit, one may write once more a first order differential equation of the form \eqref{eq:FirstOrderEffective}. Hence, at any asymptotic endpoint where \(\theta_z\to0\), the corresponding effective potential must be normalized so that \(V_{\rm eff}\) vanishes there. The two expressions \eqref{eq:EffectivePotentials-AF} 
and 
\eqref{eq:EffectivePotentials-FM}
display the central difference between the two dynamical completions:
\begin{itemize}
	\item In the antiferromagnetic model, the frequency enters quadratically. 	The effective anisotropy is
	$\alpha
		=
		\delta-\omega^2
		=
		m^2-D^2-\omega^2$ 
        in Eq.~(\ref{eq:alpha}).
	Thus the DM coupling and the internal rotation suppress the soliton window through the common combination \(D^2+\omega^2\).

	\item In the ferromagnetic model, the frequency enters linearly through the Berry phase. The fixed-frequency problem is controlled by
        $\Lambda=h-\mu\omega$ 
        in Eq.~\eqref{eq:Lambda}.
	Thus the ferromagnetic reduced problem depends on \(\delta=m^2-D^2\) 
    in Eq.~\eqref{eq:delta} and on the Berry-shifted field \(\Lambda\), but not on the antiferromagnetic combination \(D^2+\omega^2\). In particular, the sign of the shifted frequency matters.
\end{itemize}

Finally, we separate the fixed-frequency equations that determine the profiles from the physical Hamiltonian energies. After imposing the preferred pitch, the common static helical energy density is
\begin{equation}
	\mathcal E_{\rm hel}
	=
	\frac{1}{2}\theta_z^2
	+
	\frac{\delta}{2}\sin^2\theta
	+
	h\cos\theta .
\end{equation}
This is the static Hamiltonian density common to both dynamical completions.

In the antiferromagnetic model, the quadratic time derivative produces an explicit rotational contribution to the physical energy. For a rotating profile one has
\begin{equation}
	E_{\rm AF}(\omega)
	=
	\int dz\,
	\left[
	\frac{1}{2}\theta_z^2
	+
	\frac{\delta+\omega^2}{2}\sin^2\theta
	+
	h\cos\theta
	-
	\mathcal E_{\rm vac}
	\right].
\end{equation}
The profile itself, however, is determined by the effective anisotropy
$\alpha$ in Eq.~(\ref{eq:alpha}).
Thus, in the antiferromagnetic reduction, the frequency affects the energy in two ways: explicitly through the term \(\omega^2\sin^2\theta/2\), and implicitly through the profile
\(\theta=\theta_\omega(z)\).

In the ferromagnetic model the time dependence enters through the first-order Berry term. Consequently the Hamiltonian does not acquire a quadratic rotational energy. The physical energy keeps the static helical form,
\begin{equation}
	E_{\rm FM}(\omega)
	=
	\int dz\,
	\left[
	\frac{1}{2}\theta_z^2
	+
	\frac{\delta}{2}\sin^2\theta
	+
	h\cos\theta
	-
	\mathcal E_{\rm vac}
	\right].
\end{equation}
This does not mean that the ferromagnetic energy is independent of \(\omega\). The frequency enters implicitly through the profile, because the Berry term shifts the field appearing in the profile equation to 
	$\Lambda=h-\mu\omega$  
     in Eq.~\eqref{eq:Lambda}.
Thus \(E_{\rm FM}\) depends on \(\omega\) through \(\theta=\theta_\omega(z)\), even though no explicit \(\omega^2\) term appears in the Hamiltonian.

The extended pole-crossing profiles mentioned in the static sector also have rotating counterparts. In the following sections, however, we focus on the ordinary polar branches, namely the localized profiles whose meridional motion
stays inside \(0\leq\theta\leq\pi\). The extended branches cross the opposite pole and are therefore separated from the main classification; they are discussed in Appendix~\ref{app:pole-crossing}.

	\section{Antiferromagnetic charged solitons}\label{sec:antiferro-Q}
	
	We now turn to the antiferromagnetic dynamical completion of the rotating helical sector. The purpose of this section is to identify when the one-dimensional chiral magnet supports finite-energy charged textures of Q-ball or Q-kink type. The important point is that, in the antiferromagnetic model, the time dependence is governed by a conventional quadratic kinetic term.
Therefore a uniform internal rotation, $
\varphi(z,t)=-Dz+\omega t+\gamma 
$ in Eq.~(\ref{eq:rotating-helical})
does not merely shift a magnetic field parameter. Instead, it contributes to the reduced effective potential in the same way as the helical pitch selected by the DM interaction. After imposing the preferred pitch \(k=-D\), the equation for \(\theta(z)\) becomes the double sine-Gordon equation
	\begin{align}
	\theta''=
    \alpha
    \sin\theta\cos\theta
	-h\sin\theta ,
	\end{align}
    with $\alpha$ in Eq.~\eqref{eq:alpha}.
	This parameter controls the curvature of the reduced potential near the polar vacua. In the  relativistic Q-ball problem, the frequency reduces the effective mass through a term proportional to \(-\omega^2\). Here the DM coupling produces an additional reduction through \(-D^2\). Thus the relevant suppression is not caused by the rotation alone, but by the combined quantity
	\[
	D^2+\omega^2 .
	\]
	
\begin{figure}[h]
    \centering
    \includegraphics[width=0.92\textwidth]{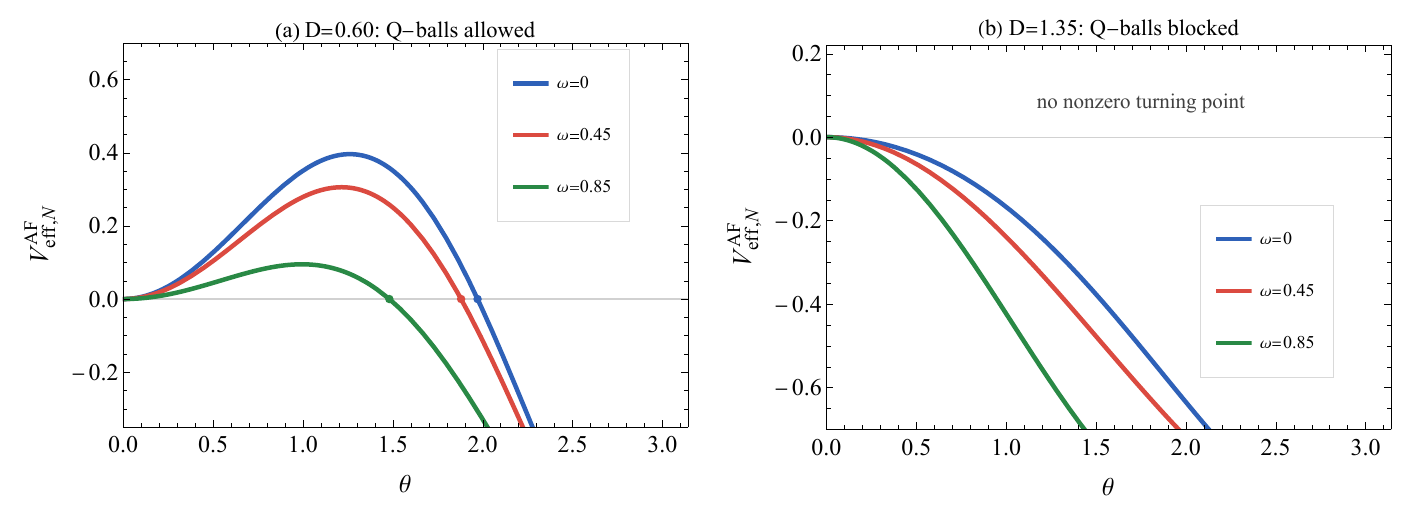}
    \caption{
    Antiferromagnetic effective potential for the north-pole Q-ball branch for \(m=\sqrt{2}\) and \(h=0.5\). Left: for \(D=0.60\), the effective potential develops a positive region and a nonzero turning point for the displayed frequencies, so localized
    Q-ball profiles exist. Right: for \(D=1.35\), the reduction of \(\alpha=\delta-\omega^2\) removes the positive region of the potential. The nonzero turning point is lost and the Q-ball branch is
    blocked.
    }
    \label{fig:af-effective-potential-allowed-blocked}
\end{figure}

	This observation is central for the existence of charged solitons. A localized Q-ball requires a nonzero turning point of the vacuum-subtracted mechanical potential, while a pole-to-pole Q-kink requires degenerate asymptotic vacua. Both requirements are affected by the same combination \(\alpha\). Consequently, increasing the helical coupling \(D\) can close the Q-soliton window even when the internal rotation frequency is small, see Fig.~\ref{fig:af-effective-potential-allowed-blocked}.
	
	In the following subsections we first analyze Q-balls localized around a polar vacuum and then the special Q-kink branch connecting the two poles. This ordering separates the two physical mechanisms: Q-balls carry Noether charge without changing the vacuum at spatial infinity, whereas Q-kinks also carry the asymptotic, topological information of the field configuration.
	
	\subsection{Antiferromagnetic Q-balls}
	
	We first consider localized charged excitations around the north pole,
	\begin{align}
	\theta(z)\longrightarrow 0,
	\qquad \mbox{at }
	z\longrightarrow \pm\infty .\label{eq:bc-AF-Q-ball}
	\end{align}
	These configurations are the antiferromagnetic Q-balls. They carry the conserved charge associated with the internal rotation, but they do not change the vacuum at spatial infinity. For this branch it is convenient to subtract the vacuum energy at \(\theta=0\). The reduced mechanical potential is then
	\begin{align}
	V^{\rm AF}_{\rm eff,N}(\theta)
	=
	\frac{\alpha}{2}\sin^2\theta
	-h(1-\cos\theta),
	\end{align}
    with $\alpha =\delta-\omega^2$ in Eq.~(\ref{eq:alpha}).
	A localized solution exists only if the mechanical particle can leave the vacuum \(\theta=0\), reach a nonzero turning point, and return to the same vacuum. This requires
	\begin{align}
	0<h<\alpha,
	\end{align}
	or equivalently
	\begin{align}
	D^2+\omega^2<m^2-h .
	\end{align}
    It is important to specify the energetic reference state of the polar Q-ball branches. For \(h>0\), the only polar Q-ball branch of this type is the north-pole branch with the boundary condition in Eq.~(\ref{eq:bc-AF-Q-ball}), 
and with the existence window \(0<h<\alpha\). Since
\begin{align}
    \mathcal E(0)=h,
    \qquad
    \mathcal E(\pi)=-h,
\end{align}
the north pole is the higher-energy polar state, while the south pole is the true homogeneous ground state. The north-pole Q-ball is therefore a finite-energy excitation relative to the metastable north-pole background,
obtained by subtracting the north-pole energy density. It is not a finite-energy excitation above the absolute ground state on the infinite line.

For \(h<0\), the roles of the two poles are reversed: the north pole is the true ground state, and the corresponding polar Q-ball branch is the south-pole branch, finite relative to the metastable south-pole background. Thus the
polar Q-balls in this minimal model are naturally interpreted as localized charged excitations in a fixed-asymptotic-sector, or metastable-background, description.

	This condition also shows that the Zeeman field imposes an absolute upper bound on the existence of the north-pole antiferromagnetic Q-ball. From Eqs.~\eqref{eq:alpha} and 
    \eqref{eq:delta}, 
	the maximum possible value of \(\alpha\) occurs at \(\omega=0\). Therefore, if $	h\geq \delta$, there is no real value of \(\omega\) for which the inequality
	\begin{align}
	0<h<\delta-\omega^2
	\end{align}
	can be satisfied. In this case the nonzero turning point is absent for every rotation frequency, and the north-pole Q-ball branch is completely removed. When \(0<h<\delta\), the allowed frequencies form the symmetric interval
	\begin{align}
	-\sqrt{\delta-h}
	<
	\omega
	<
	\sqrt{\delta-h}.
	\end{align}
	Thus the antiferromagnetic branch does not select a preferred direction of internal rotation: positive and negative frequencies are both allowed, provided their magnitude is small enough. Thus the antiferromagnetic Q-ball window is reduced jointly by the helical	DM coupling and by the internal rotation frequency. This is one of the main differences from the standard relativistic Q-ball mechanism: a sufficiently large \(D\) can remove the nonzero turning point even for small \(\omega\).

\begin{figure}[h]
    \centering
    \begin{minipage}[c]{0.58\textwidth}
        \centering
        \includegraphics[height=0.52\textwidth,keepaspectratio]{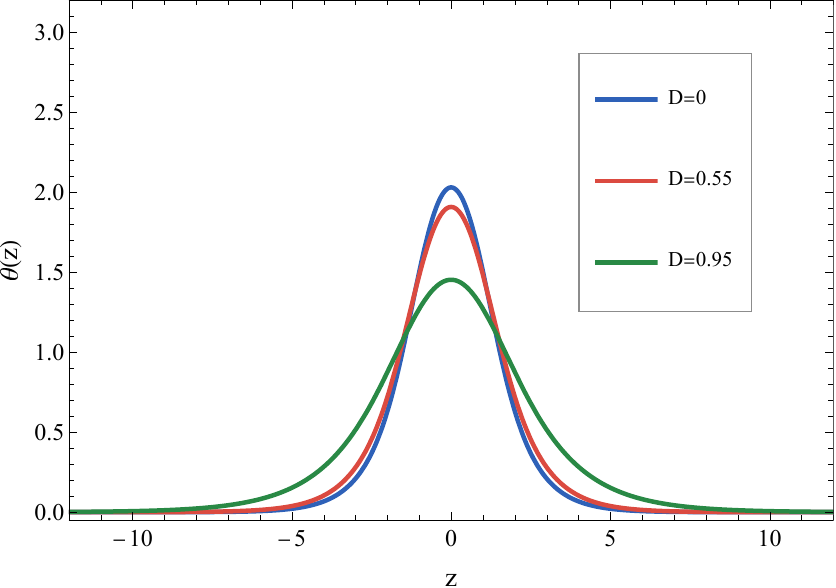}
        \par\smallskip
        (a) AF Q-ball profiles.
    \end{minipage}
    \hfill
    \begin{minipage}[c]{0.34\textwidth}
        \centering
        \includegraphics[width=\textwidth]{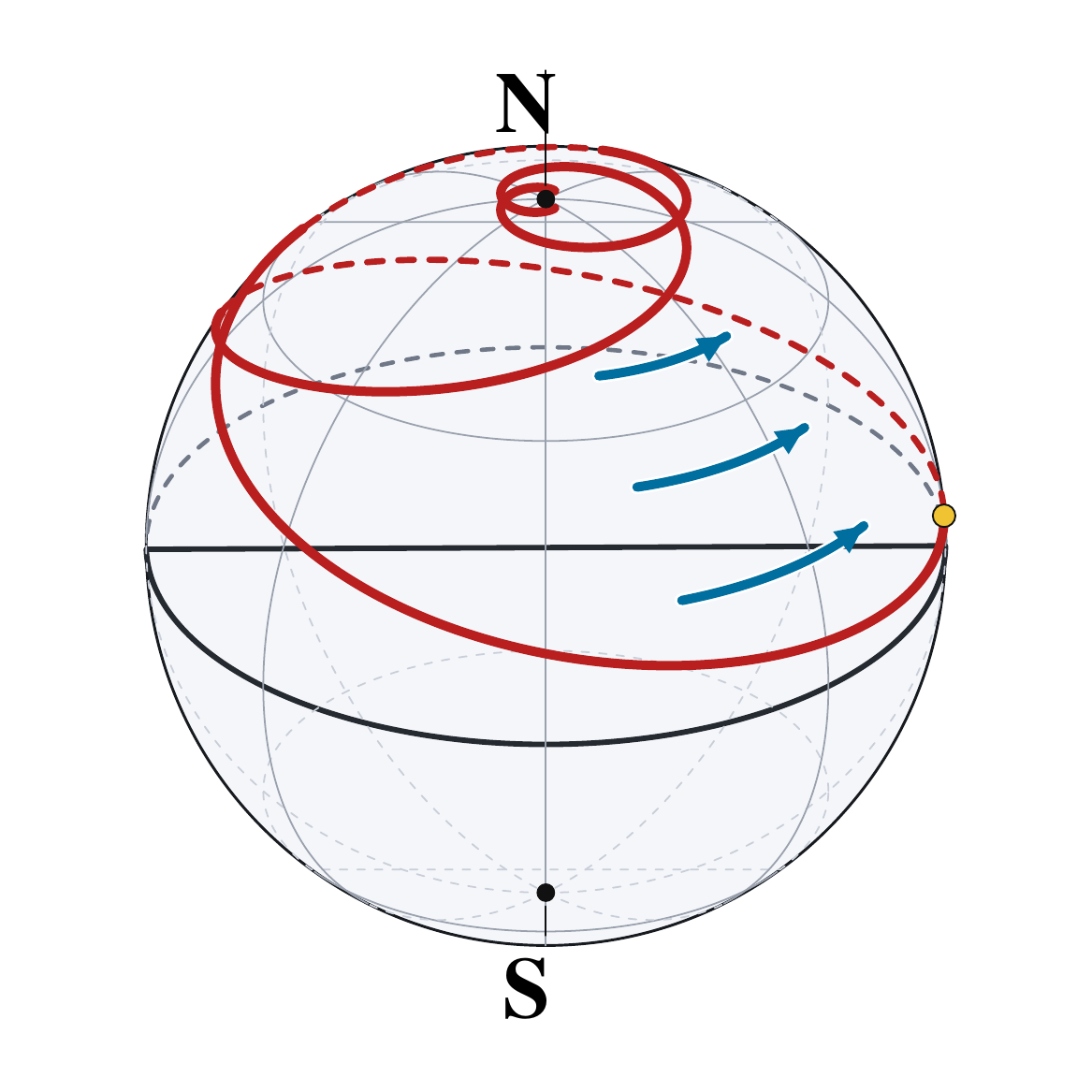}
        \par\smallskip
        (b) Rotating helical orbit on the target sphere.
    \end{minipage}
    \caption{
    Antiferromagnetic Q-ball profile and its rotating interpretation on the target sphere for the north-pole branch. Left: Q-ball profiles for several values of \(D\), with \(m=\sqrt{2}\), \(h=0.50\), and \(\omega=0.45\). Increasing \(D\) lowers \(\alpha=\delta-\omega^2\), broadens the profile, and moves the solution closer to the boundary of existence. Right: the corresponding non-topological excursion on the sphere, together with the internal rotation indicated by the azimuthal arrow.
    }
    \label{fig:af-qball-profile-sphere}
\end{figure}

	For the north-pole branch, the corresponding analytic solution for the first-order equation for the localized profile is
	\begin{align}
	\tan\frac{\theta_N(z)}{2}
	=
	\sqrt{\frac{\alpha-h}{h}}\,
	{\rm sech}
	\left[
	\sqrt{\alpha-h}\,(z-z_0)
	\right]\,,
	\end{align}
	where the parameter \(z_0\) is the center of the Q-ball, see Fig.~\ref{fig:af-qball-profile-sphere}. The solution exists only when the square roots are real, which is precisely the condition written above.

    Notice that the corresponding south-pole branch is obtained by the reflection
\begin{align}
    h\mapsto -h,
    \qquad
    \theta\mapsto \pi-\theta .
\end{align}
Thus it exists for \(h<0\) with
\begin{align}
    0<|h|<\alpha,
\end{align}
or equivalently
\begin{align}
    D^2+\omega^2<m^2-|h| .
\end{align}
In the following we display only north-pole expressions explicitly; the south-pole expressions follow from the same transformation.
	
\begin{figure}[t]
    \centering
    \includegraphics[width=\textwidth]{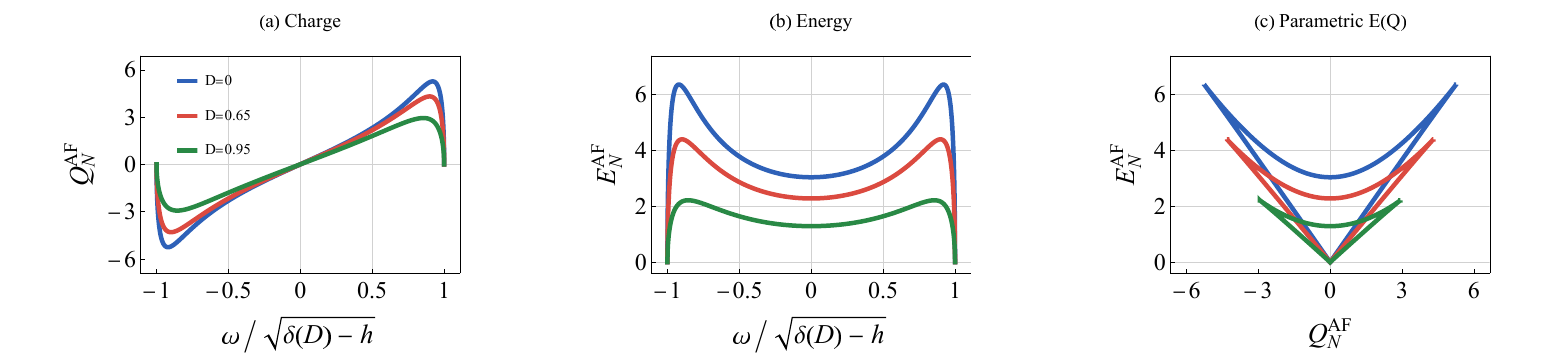}
    \caption{
    {Antiferromagnetic Q-ball for \(D=0,0.65,0.95\), \(h=0.50\), and \(m=\sqrt{2}\), so that \(\delta=2-D^2\)}. The charge \(Q_N^{\rm AF}\) is odd in \(\omega\), reaches a finite extremum, and returns to zero at the edge of the existence interval. The right panel shows the parametric energy-charge curve relevant for reduced-sector fission comparisons. The parametric \(E(Q)\) curve is two-valued because \(Q_N^{\rm AF}(\omega)\) is non-monotonic. For a fixed charge below \(Q_{\max}\), the lower-energy branch is the physically relevant fixed-charge candidate, while the upper branch is energetically disfavored.
    }
    \label{fig:af-qball-observables}
\end{figure}

	The antiferromagnetic charge is obtained from the Noether charge of the reduced \(U(1)\) rotation \eqref{eq:AF-charge-general}. For the explicit profile one finds
	\begin{align}
	Q^{\rm AF}_{N}(\omega,D)
	=
	\omega
	\left[
	\frac{4\sqrt{\alpha-h}}{\alpha}
	+
	\frac{4h}{\alpha^{3/2}}
	{\rm artanh}
	\sqrt{\frac{\alpha-h}{\alpha}}
	\right]\,.
	\end{align}
	The charge is odd in \(\omega\), so reversing the sense of the internal rotation reverses the charge, as it is shown in 
    Fig.~\ref{fig:af-qball-observables}(a).
	
	The corresponding renormalized energy is
	\begin{align}
	E^{\rm AF}_{N}(\omega,D)
	=
	\frac{4\delta\sqrt{\delta-\omega^2-h}}{\delta-\omega^2}
	+
	\frac{4h\left(2\omega^2-\delta\right)}
	{\left(\delta-\omega^2\right)^{3/2}}
	{\rm artanh}
	\sqrt{
		\frac{\delta-\omega^2-h}{\delta-\omega^2}
	}\,,
	\end{align}
	which unlike the charge, is even in \(\omega\), 
    see 
    Fig.~\ref{fig:af-qball-observables}(b).
    Consequently the antiferromagnetic Q-ball branch has a symmetric energy-charge curve,
	\begin{align}
	E^{\rm AF}_{N}(Q)=E^{\rm AF}_{N}(-Q),
	\end{align}
	see Fig.~\ref{fig:af-qball-observables}(c).
    
{ 
An important consequence is that the parametric curve cannot be interpreted as a globally single-valued function \(E(Q)\). For fixed \(D\) and \(h\), the charge \(Q_N^{\rm AF}(\omega)\) vanishes in two physically different limits. At \(\omega=0\), the rotation stops and the charge vanishes, while the profile remains a non-trivial static non-topological kink with finite energy. At the endpoint \(\alpha-h\to0\), the turning point collapses back to the vacuum, and both the charge and the vacuum-subtracted energy vanish. Hence \(Q_N^{\rm AF}(\omega)\) reaches a maximum at an intermediate frequency, producing two branches of \(E(Q)\) for charges below this maximum.

This fold is the natural place where a reduced-sector stability change may occur. In the standard Vakhitov--Kolokolov theory, such a conclusion requires more than the sign of \(dQ/d\omega\): one must also control the number of negative directions and the kernel of the constrained fluctuation operator. We do not prove those spectral assumptions here. What follows directly from the energy-charge diagram is the fixed-charge energetic ordering: the branch connected to the vacuum endpoint is lower in energy, whereas the branch connected to the static non-topological kink is energetically disfavored. Since the background can be metastable, this should be understood as a local statement within the fixed asymptotic sector, not as absolute stability against decay to the true vacuum.
}

	\subsection{Antiferromagnetic Q-kinks}
	
	We now consider charged configurations which interpolate between the two polar vacua,
	\begin{align}
	\theta(-\infty)=0,
	\qquad
	\theta(+\infty)=\pi ,
	\end{align}
	or the reverse orientation. These objects are the antiferromagnetic Q-kinks, they carry the internal \(U(1)\) charge associated with the rotation frequency \(\omega\), while also carrying the asymptotic data of a domain wall. There is, however, an important restriction. A finite-energy kink can only connect vacua with the same energy density. In the present model the fixed-frequency effective potential takes different
values at the two poles when \(h\neq0\):
\begin{align}
V_{\rm eff,AF}(0)=h+C_1,
\qquad
V_{\rm eff,AF}(\pi)=-h+C_1 .
\end{align}
	Therefore the two endpoints are degenerate only when $h=0$. For nonzero Zeeman field one may still have finite-size or boundary-supported textures, but not an isolated finite-energy pole-to-pole Q-kink. Setting \(h=0\), the antiferromagnetic equation reduces to
	\begin{align}
	\theta''
	=
	\alpha\,\sin\theta\cos\theta\,,
	\end{align}
	 where once again the relevant parameter is $\alpha
     $ in Eq.~(\ref{eq:alpha}). A Q-kink exists only if $\alpha>0$, or equivalently	$D^2+\omega^2<m^2$. Thus the same mechanism found for Q-balls appears here as well. The DM coupling and the internal rotation frequency jointly reduce the soliton window.
	
\begin{figure}[h]
    \centering
    \begin{minipage}[c]{0.58\textwidth}
        \centering
        \includegraphics[height=0.52\textwidth,keepaspectratio]{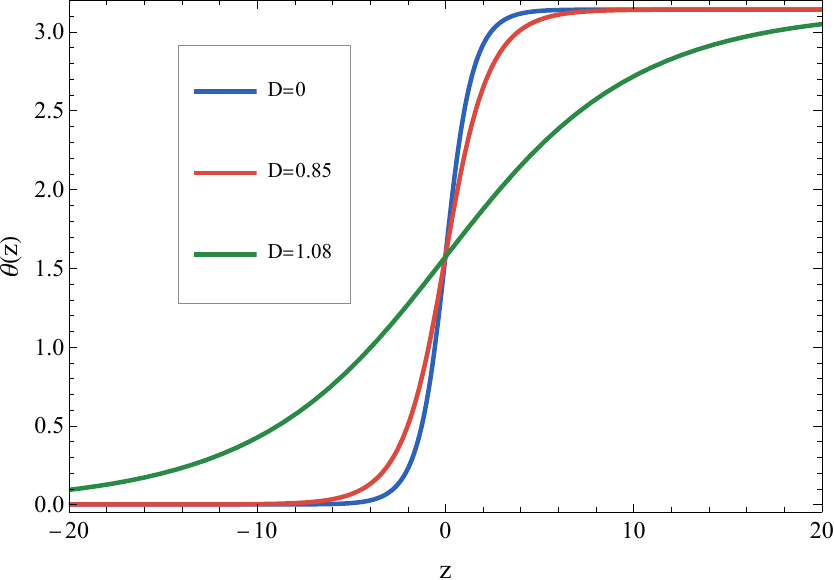}
        \par\smallskip
        (a) AF Q-kink profiles.
    \end{minipage}
    \hfill
    \begin{minipage}[c]{0.34\textwidth}
        \centering
        \includegraphics[width=\textwidth]{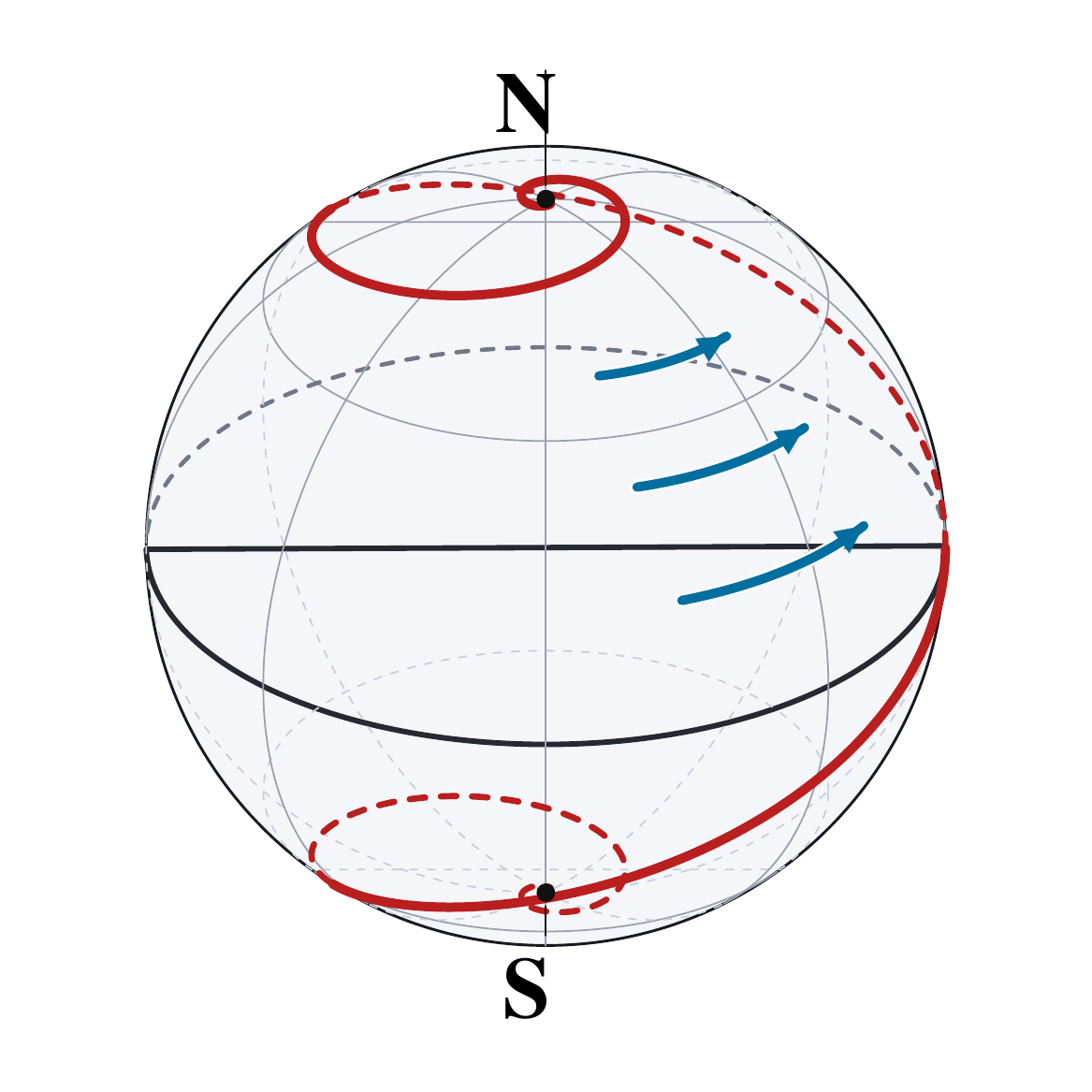}
        \par\smallskip
        (b) Rotating pole-to-pole orbit on the target sphere.
    \end{minipage}
    \caption{
    Antiferromagnetic Q-kink profile and its rotating interpretation on the target sphere. { (a): Q-kink profiles for \(D=0,0.85,1.08\), with \(h=0\), \(m=\sqrt{2}\), and \(\omega=0.90\).} In this sector the inverse width is \(\kappa=\sqrt{m^2-D^2-\omega^2}\); therefore increasing \(D\) lowers \(\kappa\), broadens the kink, and eventually destroys the localized transition when \(\kappa\) vanishes. (b): the pole-to-pole orbit on the sphere, with the internal rotation indicated by the azimuthal arrow.
    }
    \label{fig:af-qkink-profile-sphere}
\end{figure}

	The first-order equation when \(h=0\) reduces to a sine-Gordon-type equation,
with solution
\begin{align}
	\frac{1}{2}\theta'^2
	=
	\frac{\alpha}{2}\sin^2\theta ,
	\qquad
	\theta^{\rm AF}_{K}(z)
	=
	2\arctan
	{\rm e}^{
	\pm\sqrt{\alpha}\,(z-z_0)} ,
\end{align}
where \(z_0\) is the wall center and the sign distinguishes kink and antikink orientation, see Fig.~\ref{fig:af-qkink-profile-sphere}. Increasing \(D\) or \(|\omega|\) decreases the inverse width, so the kink broadens as the boundary of the existence window is approached.

We note that in the nonchiral limit $D\to0$, this solution reduces
to the precessing antiferromagnetic domain wall known in the classical
magnetic-soliton literature
\cite{KOSEVICH1990117,Baryakhtar:1979,
Baryakhtar:1980}.

\begin{figure}[h]
    \centering
    \includegraphics[width=0.32\textwidth]{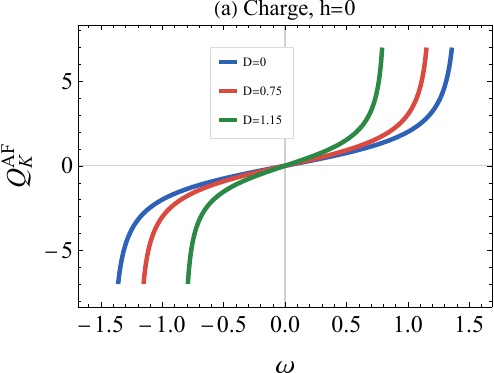}
    \hfill
    \includegraphics[width=0.32\textwidth]{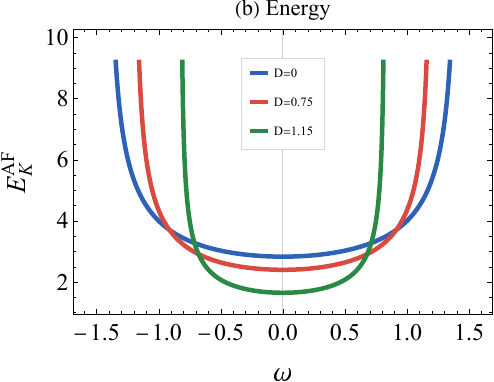}
    \hfill
    \includegraphics[width=0.32\textwidth]{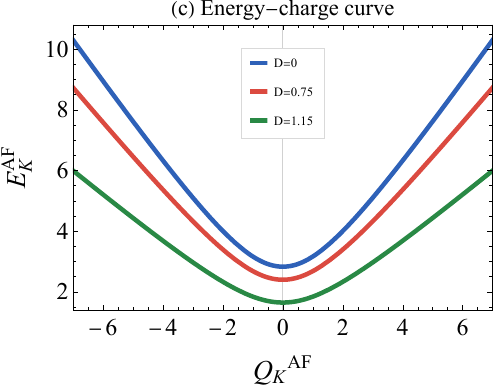}
    \caption{
    Antiferromagnetic Q-kink observables { at \(h=0\), \(m=\sqrt{2}\), and \(D=0,0.75,1.15\).} The charge is unbounded
    and diverges as \(|\omega|\to\sqrt{\delta}\). Different values of \(D\)
    change both the allowed frequency interval and the energy scale, while the
    energy-charge relation remains even in \(Q_K^{\rm AF}\).
    }
    \label{fig:af-qkink-observables}
\end{figure}

The corresponding charge and renormalized energy are
	\begin{align}
	Q^{\rm AF}_{K}(\omega,D)
	=
	\frac{2\omega}
	{\sqrt{\alpha}}\,, \qquad E^{\rm AF}_{K}(\omega,D)
	=
	\frac{2\delta}
	{\sqrt{\alpha}} ,
	\end{align}
	or eliminating the frequency, one obtains the energy-charge relation
	\begin{align}
	E^{\rm AF}_{K}(Q,D)
	=
	\sqrt{\delta}\,
	\sqrt{Q^2+4},\label{eq:AF-Q-kink-charge-energy}
	\end{align}
	see Fig.~\ref{fig:af-qkink-observables}. 
The square-root dependence on \(Q\) makes the antiferromagnetic Q-kink
energy resemble the standard relativistic charged-soliton form, although the
charge here is an internal Noether charge rather than a spatial momentum.
The expression is even in the charge, as expected for the antiferromagnetic
model: changing \(\omega\to-\omega\) reverses the Noether charge but leaves
the energy unchanged. The same energy-charge relation applies to kink and
antikink orientations.

The topology of the boundary conditions determines the relevant fission
channel. A single kink connecting the south pole to the north pole cannot
split into two isolated pole-to-pole walls while preserving the same
asymptotic vacua. The minimal nontrivial channel with the same endpoints is
an alternating three-wall configuration,
\[
    S\to N,\qquad N\to S,\qquad S\to N ,
\]
namely a kink, followed by an antikink, followed by a kink. In terms of
charged walls this channel may be written schematically as
\begin{align}
    K_{SN}(Q)
    \longrightarrow
    K_{SN}(Q_1)
    +
    \bar K_{NS}(Q_2)
    +
    K_{SN}(Q_3),
    \qquad
    Q=Q_1+Q_2+Q_3 .
\end{align}
Here \(K_{SN}\) denotes a kink from the south pole to the north pole, while
\(\bar K_{NS}\) denotes the oppositely oriented antikink. The internal charges
\(Q_i\) are determined by the corresponding rotation frequencies and are not
fixed by the topological orientation.

Using \eqref{eq:AF-Q-kink-charge-energy},
the three-wall channel satisfies
\begin{align}
    E^{\rm AF}_{K}(Q_1+Q_2+Q_3,D)
    <
    E^{\rm AF}_{K}(Q_1,D)
    +
    E^{\rm AF}_{\bar K}(Q_2,D)
    +
    E^{\rm AF}_{K}(Q_3,D),
    \qquad
    E^{\rm AF}_{\bar K}(Q_2,D)=E^{\rm AF}_{K}(Q_2,D).
\end{align}
Thus a single Q-kink carrying the total internal charge is energetically
preferred, within the reduced one-dimensional sector, over the minimal
topology-preserving kink--antikink--kink fission channel.

	This conclusion should be understood as a reduced-sector classical stability statement for the one-dimensional magnetic system. It does	not replace a full stability analysis against arbitrary perturbations of the higher-dimensional chiral magnet, but it does show that the Q-kink branch is not energetically driven to split within	the helical one-dimensional sector.
	
	\section{Ferromagnetic charged Q-balls}
    \label{sec:ferro-Q}
	
	We now consider the ferromagnetic dynamical completion of the same 
    one dimensional chiral magnet. The essential difference from the antiferromagnetic case is that the time dependence is first order. In the north-pole Berry gauge, 
in Eq.~(\ref{eq:Berry-north}),    
the frequency \(\omega\) does not enter the reduced equations through a quadratic term \(-\omega^2\). Instead, after imposing the rotating helical ansatz in Eq.~(\ref{eq:rotating-helical}),
the fixed-frequency mechanical problem is controlled by the Berry-shifted field 
\(\Lambda=h-\mu\omega\) 
in Eq.~\eqref{eq:Lambda}.
The vacuum-unsubtracted effective potential is
\begin{align}
V_{\rm eff,FM}(\theta)
=
\frac{\delta}{2}\sin^2\theta
+
\Lambda\cos\theta,
\end{align}
with $\delta$ in \eqref{eq:delta}.
	Thus the ferromagnetic frequency acts as a linear Berry-phase shift of the magnetic field. This is the central distinction from the antiferromagnetic model, where the relevant combination is	\(D^2+\omega^2\).
	
	For localized Q-balls around the north pole it is convenient to subtract the value at \(\theta=0\). This gives
	\begin{align}
	V^{\rm FM}_{\rm eff, N}(\theta)
	=
	\frac{\delta}{2}\sin^2\theta
	-
	\Lambda(1-\cos\theta).    
	\end{align}
	A nonzero turning point exists when
	\begin{align}
	0<\Lambda<\delta,
	\end{align}
	or equivalently
	\begin{align}
	0<h-\mu\omega<m^2-D^2 .
	\end{align}
	The ferromagnetic existence windows have a different interpretation because the frequency enters linearly through \(
	\Lambda=h-\mu\omega	\). For the north-pole droplet one needs
	\begin{align}
	0<h-\mu\omega<\delta,
	\end{align}
	whereas for the south-pole droplet one needs
	\begin{align}
	-\delta<h-\mu\omega<0 .
	\end{align}
	Assuming \(\mu>0\), these conditions become
	\begin{align}
	\frac{h-\delta}{\mu}<\omega<\frac{h}{\mu}
	\qquad \text{north pole},
    \end{align}
	and
	\begin{align}
	    \frac{h}{\mu}<\omega<\frac{h+\delta}{\mu}
	\qquad \text{south pole}.
	\end{align}
	Thus the two polar ferromagnetic Q-balls lie on opposite sides of the Berry-shift cancellation point value \(\omega=h/\mu\). Unlike the antiferromagnetic case, the allowed branch depends on the direction	of rotation. At fixed \(h\), one sign of the shifted rotation supports a north-pole Q-ball, while the opposite sign supports a south-pole Q-ball.
    
\begin{figure}[h]
    \centering
    \begin{minipage}[h]{0.31\textwidth}
        \centering
        \includegraphics[width=\textwidth]{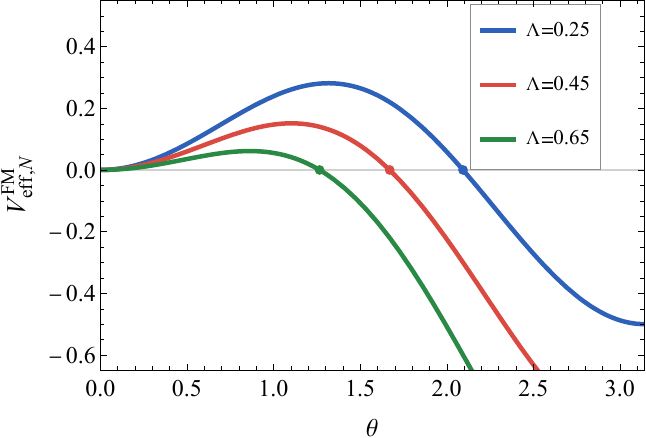}
        (a) Effective potential for different values of \(\Lambda=h-\mu\omega\).
        \label{fig:fm-potential}
    \end{minipage}
    \hfill
    \begin{minipage}[h]{0.31\textwidth}
        \centering
        \includegraphics[width=\textwidth]{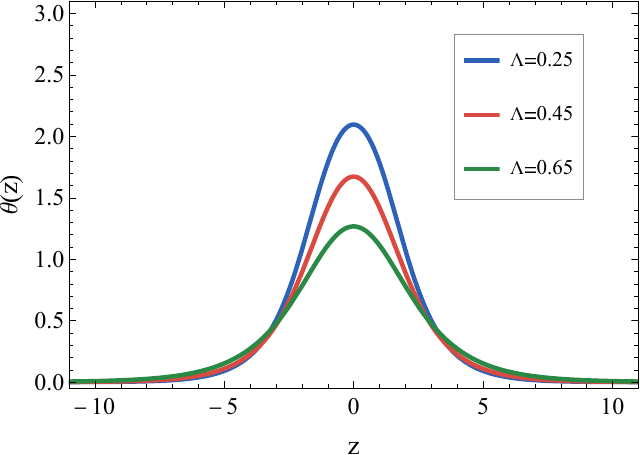}
        (b) North-pole ferromagnetic Q-ball profiles.
        \label{fig:fm-profiles}
    \end{minipage}
    \hfill
    \begin{minipage}[h]{0.31\textwidth}
        \centering
        \includegraphics[width=\textwidth]{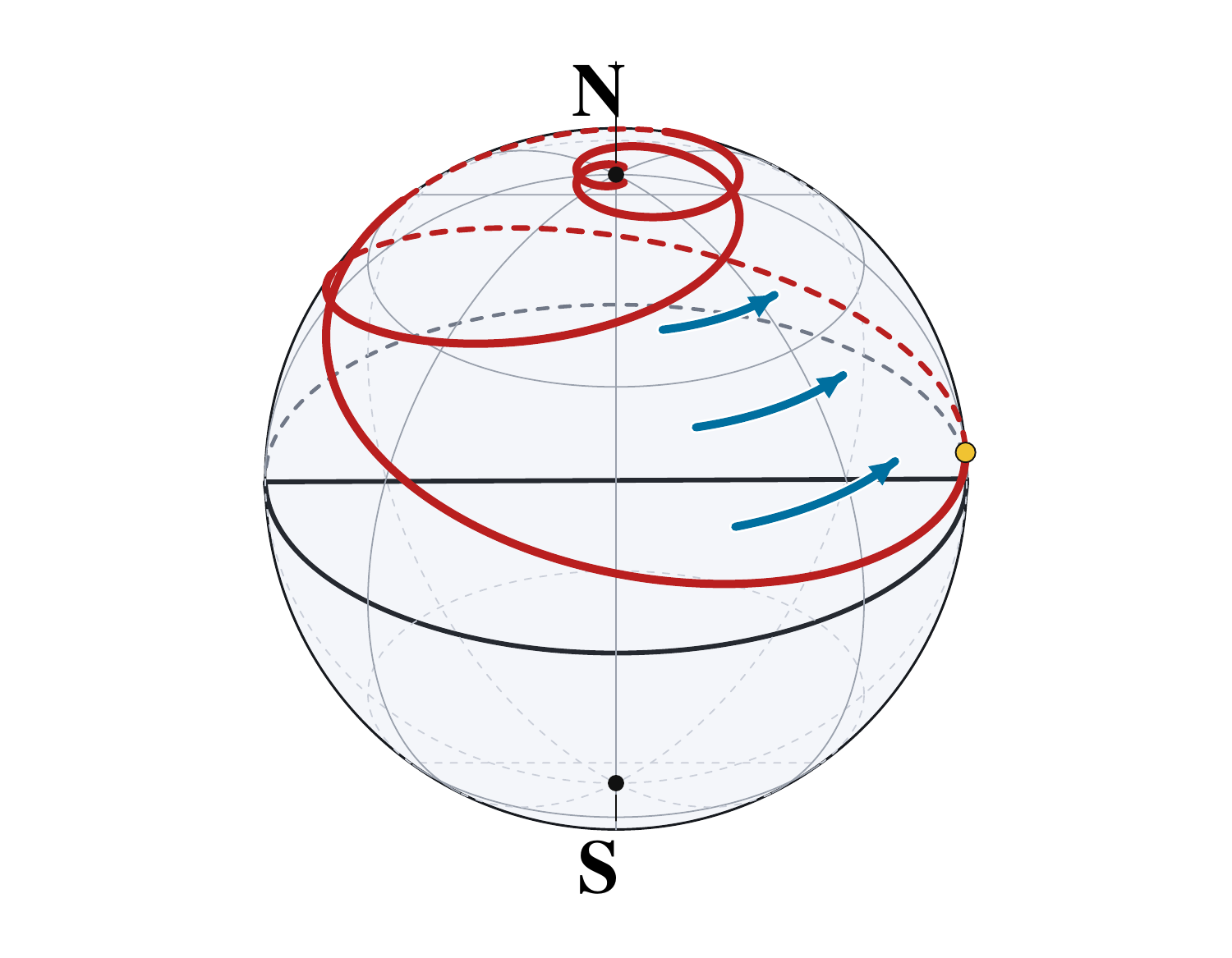}
        (c) Rotating orbit on the target sphere.
        \label{fig:fm-qball-sphere}
    \end{minipage}

    \caption{
Ferromagnetic reduced dynamics in the helical sector for the north-pole branch{, with \(\delta=1\)}. The relevant control parameter is the Berry-shifted field \(\Lambda=h-\mu\omega\). For \(0<\Lambda<\delta\), localized north-pole Q-balls exist. The profile is controlled by \(\Lambda\), while the
corresponding charge is the positive Berry-phase charge measured in the gauge regular at the north pole.}
    \label{fig:fm-reduced-dynamics}
\end{figure}

	In the north-pole window, the profile is 
	\begin{align}
	\tan\frac{\theta_N(z)}{2}
	=
	\sqrt{\frac{\delta-\Lambda}{\Lambda}}\,
	{\rm sech}
	\left[
	\sqrt{\delta-\Lambda}\,(z-z_0)
	\right]\,,
    \label{eq:ferro-Q-N}
	\end{align}
	see Fig.~\ref{fig:fm-reduced-dynamics}. The shape is therefore controlled by the Berry-shifted field \(\Lambda\), rather than by the antiferromagnetic combination \(D^2+\omega^2\). 

In the case of a small $\Lambda$, this solution can be regarded as a well-separated kink--antikink pair, with the vacuum at the opposite pole appearing between them. As $\Lambda$ approaches zero\footnote{Note that we can show that this profile satisfies 
\begin{align} 
   \forall \epsilon \in \mathbb R_{>0}, \quad &\lim_{\Lambda\to +0} \Big| \left\{ z \in \mathbb R|  0< \pi-\theta_N(z) < \epsilon  \right\}\Big|  =
   \lim_{\Lambda\to +0} \frac{1}{\sqrt{\delta }}\log \frac{\delta \epsilon^2}{4\Lambda }=\infty,
\end{align} 
and that is, $\theta_N(z)$ approaches $\pi$  as $\Lambda$ approaches zero and 
the width of the profile expands to infinity.}, the distance between them becomes infinite,  
leaving the opposite vacuum.

The south-pole profile is obtained by the reflection 
    \begin{align}
        \Lambda\mapsto-\Lambda \,, \qquad \theta\mapsto\pi-\theta\,,
        \end{align}
    with the physical south-pole window corresponding to negative shifted field \(-\delta<\Lambda<0\).
    
	The ferromagnetic Noether charge is tied to the Berry-phase one-form and therefore depends on the choice of Berry gauge. For the north-pole branch, where the field approaches \(\theta=0\) at spatial infinity, the Berry gauge regular at the north pole gives the natural vacuum-subtracted charge
\begin{align}
Q_N^{\rm FM}
=
\mu\int_{-\infty}^{\infty}
\left(1-\cos\theta_N\right)\,dz 
\end{align}
where $\theta_N(z)$ 
is a solution of Eq.~\eqref{eq:ferro-Q-N}. 
The integrand \(1-\cos\theta\) is the solid-angle density measured relative to the north pole. Analogously, for a south-pole branch one uses the Berry gauge regular at \(\theta=\pi\), replacing \(1-\cos\theta\) by \(1+\cos\theta\), the solid-angle density measured relative to the south pole. Thus the finite ferromagnetic charge is gauge-adapted to the polar background approached at spatial infinity. Because \(1-\cos\theta_N\geq0\), the north-pole charge is non-negative for \(\mu>0\), and it is strictly positive for every nontrivial north-pole Q-ball.

{
With this convention the north-pole charge is a positive vacuum-subtracted Berry charge. Along the north-pole branch one finds
\[
\frac{dE_N^{\rm FM}}{dQ_N^{\rm FM}}=-\omega .
\]
Thus the thermodynamic variable conjugate to the positive charge is \(-\omega\). A VK-type slope condition, if applied to the ferromagnetic Berry-phase dynamics, must therefore be formulated with this convention in mind and cannot be copied directly from the relativistic antiferromagnetic case. In what follows we use the monotonicity of \(Q_N^{\rm FM}(\omega)\) only to describe the branch structure; a genuine linear-stability criterion for the ferromagnetic branch would require a separate fluctuation analysis.}

\begin{figure}[h]
    \centering
    \includegraphics[width=\textwidth]{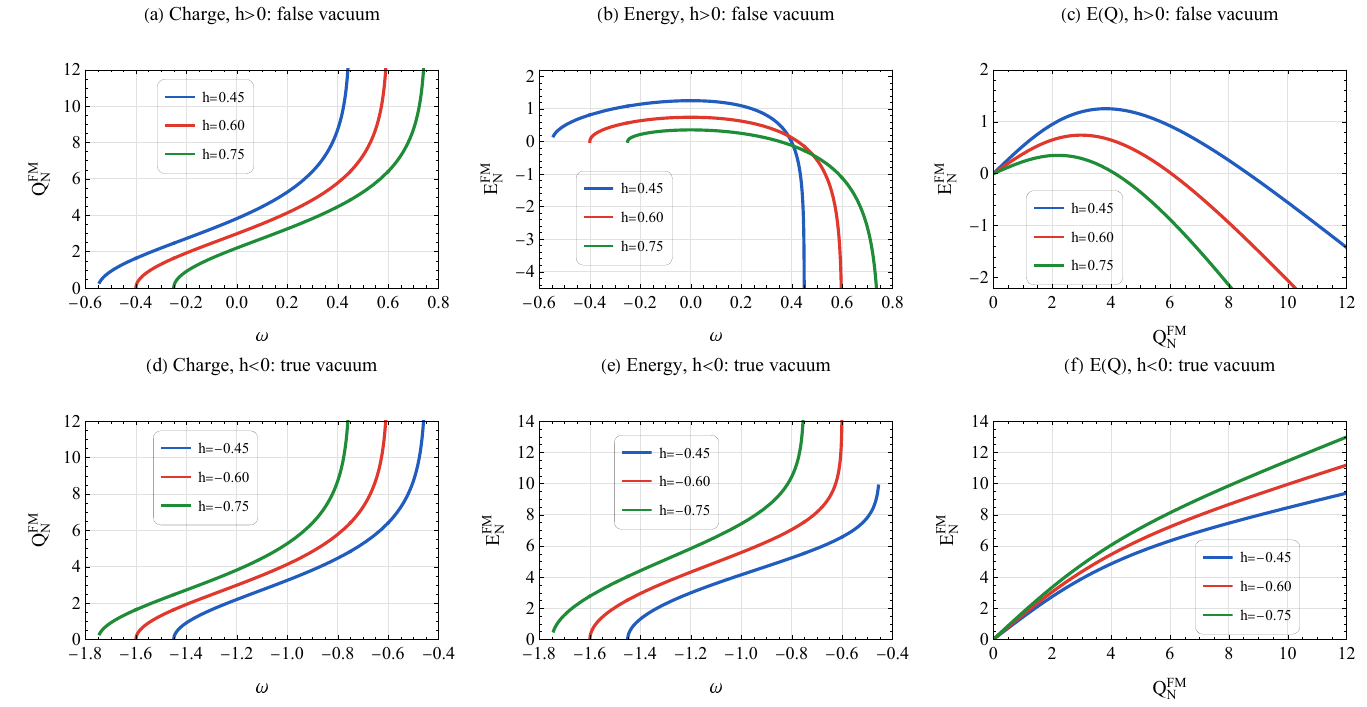}
    \caption{
    {
    Ferromagnetic Q-ball observables for different values of \(h\), with \(\mu=\delta=1\). The top row shows the case \(h>0\), for which the north pole is the false polar vacuum. The bottom row shows the case \(h<0\), for which the
north pole is the true polar vacuum. From left to right, the columns show \(Q_N^{\rm FM}(\omega)\), the vacuum-subtracted energy \(E_N^{\rm FM}(\omega)\), and the tilted energy-charge relation \(E_N^{\rm FM}(Q_N^{\rm FM})\) for \(Q_N^{\rm FM}\geq0\). Changing \(h\) shifts the allowed branch along the \(\omega\)-axis because the profile and charge depend on \(\Lambda=h-\mu\omega\).}
    }
    \label{fig:fm-qball-observables}
\end{figure}

For the explicit profile for the north-pole branch one finds
\begin{equation}\label{eq:ChargeFMQ}
    Q^{\rm FM}_{N}(\omega)
    =
    \frac{4\mu}{\sqrt{\delta}}
    {\rm artanh}
    \sqrt{
        \frac{\delta-h+\mu\omega}{\delta}
    },
\end{equation}
{ see the first column of Fig.~\ref{fig:fm-qball-observables}}, where the existence condition is \(0<h-\mu\omega<\delta
\). Equivalently, since \(\Lambda=h-\mu\omega\), the argument of the square root is \((\delta-\Lambda)/\delta\), which lies strictly between \(0\) and \(1\). Hence \(Q_N^{\rm FM}(\omega)>0\) throughout the north-pole Q-ball branch as expected. As \(\Lambda\to0^+\), the north-pole Q-ball develops an increasingly wide interior region in which \(\theta_N\simeq\pi\). On every fixed finite interval around the center, the profile therefore approaches the south-pole vacuum, which has zero charge in the south-pole Berry gauge. However, for every finite \(\Lambda>0\), the profile still returns to the north pole as \(z\to\pm\infty\). The endpoint should therefore be understood as the limit in which the south-pole domain expands without bound, rather than as a finite-charge transition between two ordinary Q-ball branches.
	
	The physical energy is obtained from the Hamiltonian, not from the fixed-frequency mechanical potential. For the north-pole branch, after subtracting the vacuum energy at \(\theta=0\), one finds
	\begin{align}
	E^{\rm FM}_{N}
	=
	4\sqrt{\delta-\Lambda}
	-
	\frac{h}{\mu}Q^{\rm FM}_{N}    
	\end{align}
     with \(\Lambda=h-\mu\omega\),
    { see the second column of Fig.~\ref{fig:fm-qball-observables}}.
	Thus the frequency controls the shape and charge through \(\Lambda\), while the Zeeman field appears explicitly in the physical energy. This is another way in which the ferromagnetic branch differs from the antiferromagnetic Q-ball: the Berry phase shifts the equations of motion, but it does not contribute directly to the Hamiltonian.
	
	It is useful to eliminate the frequency in favor of the charge. From Eq.~\eqref{eq:ChargeFMQ}, the energy-charge relation for the fixed north-pole branch can be written, for \(Q=Q_N^{\rm FM}\ge0\), as
	\begin{equation}
	E^{\rm FM}(Q)
	=
	4\sqrt{\delta}
	\tanh\left(
	\frac{\sqrt{\delta}\, Q}{4\mu}
	\right)
	-
	\frac{h}{\mu}Q ,\label{eq:FM_Energy_Charge}
	\end{equation}
	{see the third column of Fig.~\ref{fig:fm-qball-observables}}. For the fixed north-pole branch considered here, \(Q=Q_N^{\rm FM}\) is non-negative by construction. The energy-charge relation should therefore be read as a one-branch relation for \(Q\ge0\), rather than as a symmetric two-sided curve.  The sign of the Zeeman field determines the tilt of this branch: for \(h>0\), the term \(-hQ/\mu\) lowers the renormalized energy as the positive charge increases, whereas for \(h<0\) the same term raises the energy. Thus the two rows of Fig.~\ref{fig:fm-qball-observables} represent the same monotonic charge branch shifted in \(\omega\), but with opposite energetic tilts.
 
{ The decrease of the vacuum-subtracted energy for \(h>0\) is not a contradiction. It reflects that the subtraction is made relative to the metastable north-pole background. As \(Q_N^{\rm FM}\) increases, the Q-ball develops an increasingly wide interior region close to the lower-energy south-pole state. Thus fixed-charge comparisons along the branch remain meaningful, while unconstrained minimization drives the solution toward the \(\Lambda\to0^+\) endpoint of the reduced description.}

For \(-\delta<\Lambda<0\), the south-pole branch has
\[
Q_S^{\rm FM}(\Lambda)
=
\frac{4\mu}{\sqrt{\delta}}\,
{\rm arctanh}
\sqrt{\frac{\delta+\Lambda}{\delta}},
\qquad
E_S^{\rm FM}(\Lambda)
=
4\sqrt{\delta+\Lambda}
+\frac{h}{\mu}Q_S^{\rm FM}.
\]
Thus the two branches have the same nonlinear dependence on
\(|\Lambda|\), but the Zeeman contribution enters with opposite sign for the
two polar backgrounds.

    The fixed-pole ferromagnetic branch has a simpler branch structure than the antiferromagnetic Q-ball family. Since \(Q_N^{\rm FM}(\omega)\) is monotonic on the interval \(0<h-\mu\omega<\delta\), a fixed positive charge selects a unique north-pole profile. However, the renormalised energy is not bounded from below as a function of \(Q\) when \(h>0\). Indeed, \(Q_N^{\rm FM}\to\infty\) as \(h-\mu\omega\to0^+\), while the energy-charge relation \eqref{eq:FM_Energy_Charge} is eventually dominated by the negative linear Zeeman term. The finite critical point of \(E_N^{\rm FM}(Q)\) is therefore a local maximum, not a minimum:
\begin{align}
Q_{\rm max}^{E}
=
\frac{4\mu}{\sqrt{\delta}}
\operatorname{artanh}
\sqrt{1-\frac{h}{\delta}} .
\end{align}
Thus the negative part of the curve should not be interpreted as a stable finite-size optimum. It indicates that, relative to the chosen vacuum subtraction, increasing the droplet charge becomes energetically favorable as the branch approaches the \(\Lambda=0\) endpoint.

The negative renormalized energy at large \(Q_N^{\rm FM}\) should not be read as a stable finite-charge minimum. It reflects the fact that, for \(h>0\), the north-pole background is the higher-energy polar state. As the branch approaches \(\Lambda\to0^+\), the droplet charge diverges and the configuration increasingly resembles a conversion of the north-pole background toward the lower-energy south-pole state. Thus the branch is meaningful at fixed charge, while unconstrained minimization drives the solution toward the endpoint of the reduced description.

	For the fixed north-pole branch, the reduced-sector fission comparison should be made at positive charge. Within the same one-dimensional reduced sector, for \(Q_1,Q_2>0\)
\begin{align}
E_N^{\rm FM}(Q_1+Q_2)
<
E_N^{\rm FM}(Q_1)+E_N^{\rm FM}(Q_2) \,.
\end{align}

	Thus, a single ferromagnetic charged droplet carrying the total charge is energetically preferred over two separated droplets with the same total charge. This is the ferromagnetic analogue of the fission-stability statement obtained for the antiferromagnetic branch, but its origin is different. In the antiferromagnetic model the charge dependence comes from the relativistic quadratic frequency term, whereas here it follows from the Berry-phase relation between the charge and the shifted Zeeman parameter \(\Lambda=h-\mu\omega\). As before, this is a classical stability statement for the one-dimensional chiral magnets, but not a proof of stability against arbitrary perturbations of the higher dimensional chiral magnets.

	The endpoint \(\Lambda=0\), which lies at the boundary of the Q-ball window shown in Fig.~\ref{fig:fm-reduced-dynamics}, requires separate interpretation. In the reduced fixed-frequency mechanical problem one then obtains
\begin{align}
V_{\rm eff,FM}^{(\Lambda=0)}(\theta)
=
\frac{\delta}{2}\sin^2\theta
+
C_2 .
\end{align}
Choosing \(C_2=0\), the first-order equation admits the formal pole-to-pole
solution
	\begin{align}
	\theta(z)
	=
	2\arctan
	\exp\left[
	\pm\sqrt{\delta}\,(z-z_0)
	\right].
	\end{align}
	This solution should be interpreted with some care. The condition
\(\Lambda=0\) makes the two poles degenerate in the fixed-frequency mechanical problem, but it does not by itself make them degenerate in the physical Hamiltonian. The Berry term changes the equations of motion and shifts the effective mechanical tilt, but it does not contribute to the laboratory-frame energy. The physical static energy densities at the two poles remain
\begin{align}
    \mathcal E(0)=h,
    \qquad
    \mathcal E(\pi)=-h .
\end{align}
Therefore, when \(h\neq0\), no single vacuum subtraction can make the energy
density vanish at both spatial infinities. A pole-to-pole configuration at
\(\Lambda=0\) then has infinite laboratory-frame energy on the infinite line unless \(h=0\).

The special case \(h=0\) is different but should not be called a charged
ferromagnetic Q-kink. In that case \(\Lambda=0\) implies \(\omega=0\), and the solution reduces to the ordinary static domain wall of the magnetic energy. This wall has finite energy because the two polar vacua are physically degenerate. However, it does not carry a finite localized ferromagnetic Noether charge relative to a single polar Berry gauge: in the north-pole gauge the charge density approaches a nonzero constant on the south-pole side, while in the south-pole gauge the analogous divergence appears on the north-pole side. Thus the divergence reflects the change of asymptotic background, not a localized infinite charge on the wall. Consequently, the ferromagnetic model has a formal pole-to-pole solution at \(\Lambda=0\), and a finite-energy static wall when \(h=0\), but it does not provide an isolated finite-energy, finite-charge ferromagnetic Q-kink on the infinite line in this minimal setting.

Geometrically, this reflects the fact that the ferromagnetic Berry charge is defined relative to a polar gauge. The north-pole density \(1-\cos\theta\) and the south-pole density \(1+\cos\theta\) are the solid-angle densities measured from opposite poles. A pole-to-pole wall changes the asymptotic pole, so no single Berry gauge, or single solid-angle subtraction, makes the charge density vanish at both spatial infinities.

\section{Comparison of the mechanisms}\label{sec:comparison}


The antiferromagnetic and ferromagnetic theories considered above are two
dynamical completions of the same static chiral magnetic functional. They
therefore share the same helical static sector, but they convert an internal
rotation into a fixed-frequency profile problem in fundamentally different
ways.

In the polar finite-energy sector studied in this work, the helical ansatz
follows the DM-preferred pitch and carries no longitudinal phase current,
\(j^z=0\). Sectors with \(j^z\neq0\) are qualitatively different, since the
elimination of \(\varphi_z\) produces a centrifugal term singular at the polar
vacua. The comparison below is therefore a comparison within the polar helical
sector.

\begin{figure}[h]
    \centering
    \includegraphics[width=\textwidth]{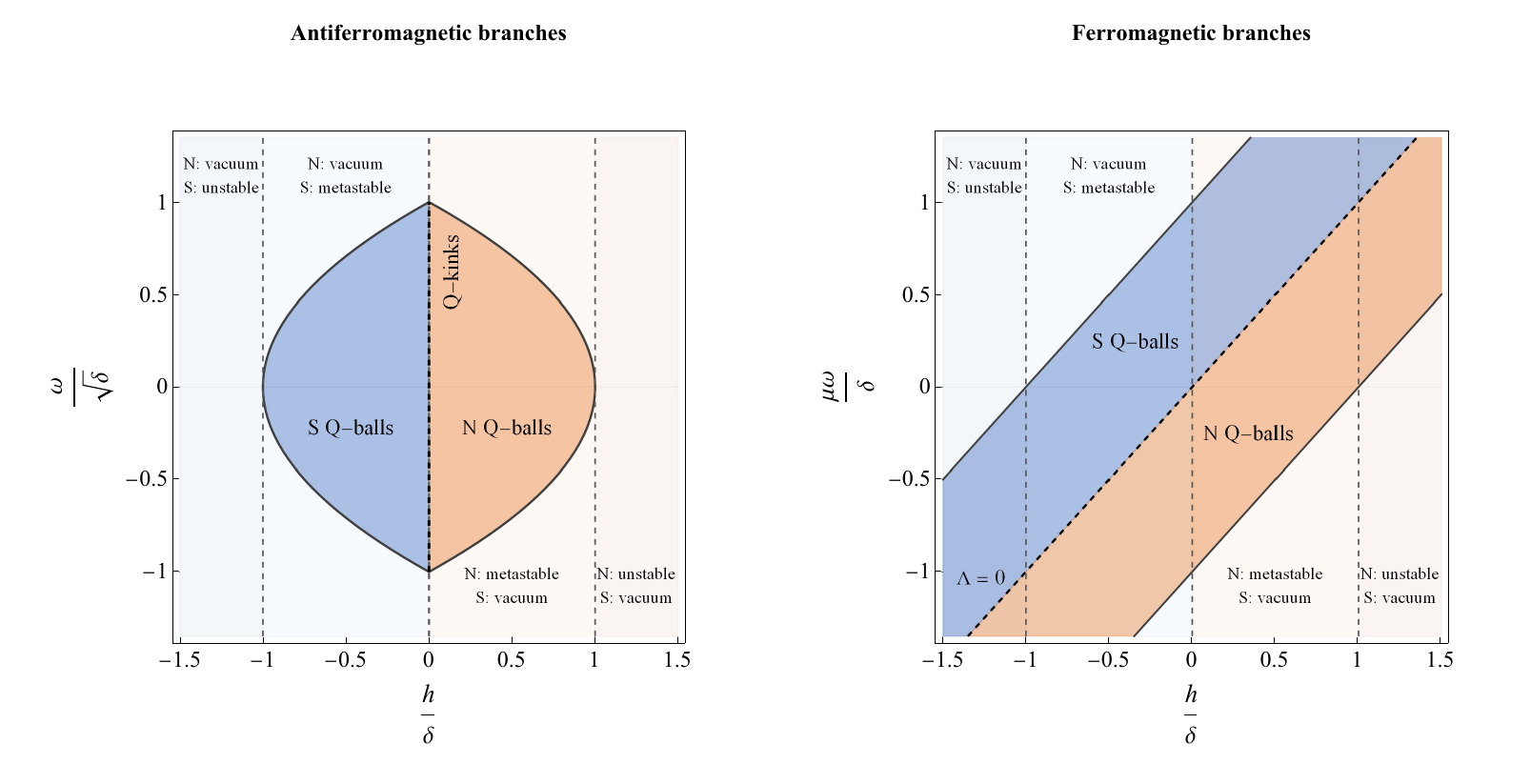}
    \caption{
        Existence regions for charged helical solitons in normalized variables.
        Left: antiferromagnetic branches, with horizontal axis \(h/\delta\)
        and vertical axis \(\omega/\sqrt{\delta}\). The Q-ball windows are
        controlled by \(\alpha=\delta-\omega^2\), where
        \(\delta=m^2-D^2\), and finite Q-kinks occur on the degenerate line
        \(h=0\).
        Right: ferromagnetic branches, with horizontal axis \(h/\delta\)
        and vertical axis \(\mu\omega/\delta\). The Q-ball windows are
        controlled by the Berry-shifted field
        \(\Lambda=h-\mu\omega\), so the north- and south-pole branches occupy
        the diagonal strips \(0<\Lambda<\delta\) and
        \(-\delta<\Lambda<0\), respectively.
        The vertical lines \(h/\delta=-1,0,1\) distinguish the character of
        the polar extrema of the static Hamiltonian: true stable, false stable,
        or linearly unstable.  Only the ordinary polar Q-ball and Q-kink branches are shown; the extended
        meridional branches are discussed in Appendix~\ref{app:pole-crossing}.
    }
    \label{fig:existence-regions}
\end{figure}

In the antiferromagnetic model the time derivative enters quadratically. The
reduced Q-ball problem is controlled by
\[
    \alpha=m^2-D^2-\omega^2
    =\delta-\omega^2 .
\]
Thus the DM coupling and the rotation frequency suppress the Q-ball window through the same combination, \(D^2+\omega^2\). For the north-pole branch the existence condition is \(0<h<\alpha\), while the south-pole branch is obtained by \(h\mapsto -h\). Hence, for \(h\neq0\), the antiferromagnetic polar Q-ball is always built over the metastable, higher-energy polar background. The allowed frequency window is symmetric under \(\omega\mapsto-\omega\), { see the left panel of Fig.~\ref{fig:existence-regions}}.

The ferromagnetic model is different because the Berry phase is first order in time. The frequency does not enter through \(-\omega^2\), but through the Berry-shifted field
\[
    \Lambda=h-\mu\omega .
\]
The north- and south-pole Q-balls are selected by opposite signs of this quantity:
\[
    0<\Lambda<\delta,
    \qquad
    -\delta<\Lambda<0 .
\]
Equivalently, for \(\mu>0\), the two branches lie on opposite sides of
\[
    \omega=\frac{h}{\mu}.
\]
This value marks the point at which the imposed internal rotation cancels the Zeeman tilt in the fixed-frequency mechanical problem. Unlike in the antiferromagnetic case, the physical true and false polar vacua are selected by \(h\), whereas the ferromagnetic Q-ball branch is selected by \(\Lambda\). Thus, at fixed \(h\), varying \(\omega\) can move the system from a north-pole Q-ball to a south-pole Q-ball, { see the right panel of Fig.~\ref{fig:existence-regions}}.  The diagram summarizes the ordinary polar branches, extended meridional branches are discussed in Appendix~\ref{app:pole-crossing}.

The charge-energy relations encode the same distinction, but their
interpretation is different in the two models. In the antiferromagnetic case the Noether charge is proportional to the rotation frequency, \(Q^{\rm AF}\propto\omega\). Reversing the internal rotation therefore changes the sign of the charge while leaving the energy unchanged, so the antiferromagnetic energy-charge curves are even in \(Q\).

In the ferromagnetic case the charge is not proportional to \(\omega\). It is
the Berry, or solid-angle, charge measured relative to the asymptotic polar
background. For a north-pole Q-ball one naturally uses the Berry gauge regular at \(\theta=0\), giving a non-negative gauge-adapted charge \(Q_N^{\rm FM}\geq0\). For a south-pole Q-ball one uses the gauge regular at
\(\theta=\pi\), again obtaining a non-negative gauge-adapted charge \(Q_S^{\rm FM}\geq0\), for \(\mu>0\). Thus the two FM polar branches should not be interpreted as positive- and negative-charge sectors of a single symmetric \(E(Q)\) curve. They are two fixed-pole branches, selected by the sign of \(\Lambda=h-\mu\omega\), and each branch has its own natural vacuum subtraction and Berry-charge convention.

The comparison is especially sharp at the pole-to-pole boundary. In the
antiferromagnetic model, a Q-kink exists on the line \(h=0\), where the two
polar vacua are degenerate in the physical Hamiltonian. The same condition
also makes the fixed-frequency mechanical endpoints degenerate, so the
reduced pole-to-pole solution is a genuine finite-energy kink in the helical
sector.

In the ferromagnetic model the analogous boundary of the Q-ball windows is \(\Lambda=0\), or \(\omega=h/\mu\). At this value the imposed internal rotation cancels the Zeeman tilt in the fixed-frequency profile equation, so the reduced mechanical problem again has degenerate polar endpoints. However, this degeneracy is not a degeneracy of the physical Hamiltonian: the Berry term fixes the precessional dynamics but does not contribute to the static energy. Thus, unless \(h=0\), the resulting pole-to-pole profile is only a formal solution of the fixed-frequency reduced problem, not an isolated finite-energy ferromagnetic Q-kink on the infinite line.

The comparison therefore isolates three robust signatures of the two dynamics. First, antiferromagnetic charged solitons are controlled by the quadratic combination \(D^2+\omega^2\), whereas ferromagnetic droplets are controlled by the shifted field \(h-\mu\omega\). Second, in the antiferromagnetic model reversing the rotation frequency reverses the Noether charge, while in the ferromagnetic model changing the rotation frequency changes the Berry-shifted field and can switch the allowed Q-ball from one polar branch to the other; it does not produce charge reversal within a fixed polar branch. Third, antiferromagnetic polar Q-balls are necessarily attached to the metastable polar background selected by \(h\). Ferromagnetic polar Q-balls are different: because their branch is selected by the sign of \(h-\mu\omega\), they may be built over either the true or the false polar background, depending on the rotation frequency. These differences provide the practical diagnostic separating relativistic-type antiferromagnetic Q-solitons from Berry-phase ferromagnetic Q-balls.
\section{Discussion and outlook 
}
\label{sec:summary}

We have shown that the same one-dimensional chiral magnetic energy supports different charged-soliton mechanisms depending on whether it is completed by antiferromagnetic or ferromagnetic spin dynamics. In both cases the DM
interaction fixes the preferred helical pitch and lowers the effective easy-axis coefficient to \(\delta=m^2-D^2\). The charged sectors, however, are organized by different dynamical parameters: in the antiferromagnetic model the relevant combination is \(\alpha=\delta-\omega^2=m^2-D^2-\omega^2\), whereas in the ferromagnetic model it is the Berry-shifted field \(\Lambda=h-\mu\omega\). Thus, in the antiferromagnetic rotating sector, both the DM interaction and the internal rotation narrow the localized-solution window.

This distinction has a direct physical consequence. In the antiferromagnetic case, polar Q-balls exist only on the metastable polar background selected by the magnetic field: for \(h>0\) the finite-energy north-pole Q-ball is built over the false vacuum, and the south-pole branch is obtained by reversing the sign of \(h\). By contrast, in the ferromagnetic case the Q-ball branch is selected by \(\Lambda\), while the true and false polar vacua are selected by \(h\). Thus, at fixed magnetic field, changing the rotation frequency can move the localized branch between the north and south poles. Ferromagnetic Q-balls are therefore not restricted to the false vacuum; depending on \(\omega\), they may be built over either the true or the false polar background.

The conserved charges also have different meanings in the two theories. In the antiferromagnetic model the Noether charge is proportional to the rotation frequency, so reversing \(\omega\) reverses the charge while leaving the energy
unchanged. The non-monotonicity of \(Q^{\rm AF}(\omega)\) implies that the parametric \(E(Q)\) curve has two local branches at fixed charge; the branch connected to the vacuum gives the relevant fixed-charge candidate, whereas the
branch connected to the neutral static profile is energetically disfavored. For antiferromagnetic Q-kinks, the reduced energy-charge relation is subadditive, so a single charged kink is favored against fission into smaller charged kinks
within the one-dimensional helical sector.

In the ferromagnetic model the charge is instead a Berry, or solid-angle, charge defined relative to the asymptotic pole. Consequently the north- and south-pole branches have separate natural charge conventions rather than forming two signs of a single symmetric \(E(Q)\) curve. Negative vacuum-subtracted ferromagnetic
energies at large charge should therefore be read as conversion of a higher-energy polar background toward the lower-energy pole, not as an unconstrained finite-charge minimum.  In particular, staticity does not imply neutrality in the ferromagnetic theory: a nontrivial localized same-pole profile already carries vacuum-subtracted Berry charge, whereas static antiferromagnetic profiles have zero Noether charge.

The pole-to-pole sector further separates the two mechanisms. Antiferromagnetic Q-kinks require degenerate physical polar vacua and therefore occur only at \(h=0\) in the present model. In the ferromagnetic case the formal line
\(\Lambda=0\) makes the fixed-frequency mechanical endpoints degenerate, but it does not in general make the physical Hamiltonian degenerate. Unless \(h=0\), the corresponding pole-to-pole orbit is therefore a rotating-frame solution,
not an isolated finite-energy ferromagnetic Q-kink on the infinite line.

 {These results also delimit the main directions in which the present analysis should be extended.} 
 
 The polar Q-ball and Q-kink branches constructed here are charged excitations
of the zero-longitudinal-current helical branch selected by the static energy
minimization. The condition \(J^z=0\) fixes the spatial pitch to the
DM-preferred value, while the internal \(U(1)\) charge is generated by the
uniform time rotation. More general time-dependent perturbations may generate
local longitudinal currents and should be included in a full stability analysis
beyond the fixed-pitch ansatz.

The solutions constructed in this work are exact stationary solutions of the conservative dynamics. In real magnetic materials, Gilbert damping breaks the conservation of the corresponding charge and can make the frequency and profile
evolve in time. This does not invalidate the conservative solutions, but it means that their experimental realization likely requires either sufficiently weak damping on the relevant time scale or external driving, such as spin torque,
to compensate losses. This provides a natural connection, and also a sharp distinction, with driven-dissipative magnetic Q-balls \cite{ahlberg2024magnetic,Jiang:2024}. A systematic extension to Landau--Lifshitz--Gilbert dynamics, transverse stability, finite geometries, more general anisotropies, and ferrimagnetic dynamics combining Berry-phase
terms with sigma-model-like second-order kinetic terms remains an important direction for future work.

\section*{Acknowledgements}

This work is supported in part by grant PID2023-148409NB-I00 MTM funded by MCIN/AEI/10.13039/ 
501100011033 (AJBS), 
JSPS KAKENHI [Grants  No.~JP22H01221, 
  JP23K22492] (M.~N.), 
  and the WPI program
``Sustainability with Knotted Chiral Meta Matter (WPI-SKCM$^2$)'' at
Hiroshima University (M.~N.).


    \appendix

\section{Higher dimensional chiral magnets and longitudinal reduction}
\label{app:bulk-model}

In this appendix, we record the higher-dimensional chiral magnetic model from which the 
quasi-one-dimensional theory used in the main text can be obtained. We work in \(3+1\)-dimensional Minkowski space and use cylindrical coordinates \((r,\phi,z)\) on physical space 
with the metric is
	$ds^2
	=
	dt^2-dr^2-r^2d\phi^2-dz^2$.
The bulk DM interaction is
\begin{equation}
	\mathcal D
	=
	\mathbf n\cdot(\nabla\times\mathbf n).
\end{equation}
In the cylindrical coordinates it takes the form
\begin{align}
	\mathcal D
	=&\;
	\sin(\varphi-\phi)\,\theta_r
	-
	\frac{\cos(\varphi-\phi)}{r}\,\theta_\phi
	\nonumber\\
	&+
	\sin\theta\cos\theta
	\left[
	\cos(\varphi-\phi)\,\varphi_r
	+
	\frac{\sin(\varphi-\phi)}{r}\,\varphi_\phi
	\right]
	-
	\sin^2\theta\,\varphi_z .
\end{align}
This expression shows explicitly how the DM interaction couples the spatial orientation of the texture to the internal azimuthal angle.

The antiferromagnetic dynamical completion is the nonlinear sigma model on \(S^2\),
\begin{align}
	\mathcal L_{\rm AF}^{3+1}
	=&\;
	\frac{1}{2}
	\left(
	\theta_t^2+\sin^2\theta\,\varphi_t^2
	\right)
	-
	\frac{1}{2}
	\left[
	\theta_r^2+\frac{\theta_\phi^2}{r^2}+\theta_z^2
	+
	\sin^2\theta
	\left(
	\varphi_r^2+\frac{\varphi_\phi^2}{r^2}+\varphi_z^2
	\right)
	\right]+
	D\mathcal D
	-
	V(\theta).
\end{align}
Its Hamiltonian density is
\begin{align}
	\mathcal H_{\rm AF}^{3+1}
	=&\;
	\frac{1}{2}
	\left(
	\theta_t^2+\sin^2\theta\,\varphi_t^2
	\right)
	+
	\frac{1}{2}
	\left[
	\theta_r^2+\frac{\theta_\phi^2}{r^2}+\theta_z^2
	+
	\sin^2\theta
	\left(
	\varphi_r^2+\frac{\varphi_\phi^2}{r^2}+\varphi_z^2
	\right)
	\right]-
	D\mathcal D
	+
	V(\theta).
\end{align}

The ferromagnetic dynamical completion replaces the quadratic kinetic term by a Berry phase. With the north-pole and south-pole gauges,
\begin{equation}
	\mathcal L_B^{(N)}
	=
	\mu(\cos\theta-1)\varphi_t ,
\quad
    \mathcal L_B^{(S)}
	=
	\mu(\cos\theta+1)\varphi_t ,
\end{equation}
respectively,
the ferromagnetic Lagrangian density is
\begin{align}
	\mathcal L_{\rm FM}^{3+1,(N/S)}
	=&\;
	\mu(\cos\theta\mp1)\varphi_t
	-
	\frac{1}{2}
	\left[
	\theta_r^2+\frac{\theta_\phi^2}{r^2}+\theta_z^2
	+
	\sin^2\theta
	\left(
	\varphi_r^2+\frac{\varphi_\phi^2}{r^2}+\varphi_z^2
	\right)
	\right]+
	D\mathcal D
	-
	V(\theta).
\end{align}
The two Berry gauges differ by a total derivative 
$
	\mathcal L_B^{(S)}-\mathcal L_B^{(N)}
	=
	2\mu\varphi_t$ 
    as in Eq.~(\ref{eq:gauges}).
They give the same local equations of motion, but the associated vacuum-subtracted charges depend on the gauge adapted to the asymptotic pole.

Since the Berry term is first order in time, it fixes the symplectic structure but does not contribute directly to the physical Hamiltonian. Thus
\begin{align}
	\mathcal H_{\rm FM}^{3+1}
	=&\;
	\frac{1}{2}
	\left[
	\theta_r^2+\frac{\theta_\phi^2}{r^2}+\theta_z^2
	+
	\sin^2\theta
	\left(
	\varphi_r^2+\frac{\varphi_\phi^2}{r^2}+\varphi_z^2
	\right)
	\right]-
	D\mathcal D
	+
	V(\theta).
\end{align}

The full actions are
\begin{equation}
	S_{\rm AF}
	=
	\int dt\,dr\,d\phi\,dz\; r\,\mathcal L_{\rm AF}^{3+1},
	\qquad
	S_{\rm FM}
	=
	\int dt\,dr\,d\phi\,dz\; r\,\mathcal L_{\rm FM}^{3+1}.
\end{equation}
Throughout the paper we use the easy-axis plus Zeeman potential
\begin{equation}
	V(\theta)
	=
	m^2\sin^2\theta+h\cos\theta .
\end{equation}

The Euler--Lagrange equation for a field \(q=\theta,\varphi\) in cylindrical coordinates is
\begin{equation}
	\partial_t
	\frac{\partial\mathcal L}{\partial q_t}
	+
	\frac{1}{r}\partial_r
	\left(
	r\frac{\partial\mathcal L}{\partial q_r}
	\right)
	+
	\partial_\phi
	\frac{\partial\mathcal L}{\partial q_\phi}
	+
	\partial_z
	\frac{\partial\mathcal L}{\partial q_z}
	-
	\frac{\partial\mathcal L}{\partial q}
	=
	0 .
\end{equation}

To obtain the 
quasi-one-dimensional model used in the main text, we restrict to longitudinal configurations,
\begin{equation}
	\theta=\theta(t,z),
	\qquad
	\varphi=\varphi(t,z),
\end{equation}
so that all \(r\)- and \(\phi\)-derivatives vanish. In this sector the DM density reduces to
\begin{equation}
	\mathcal D
	=
	-\sin^2\theta\,\varphi_z .
\end{equation}
Working per unit transverse area, or equivalently absorbing the transverse volume factor into the normalization of the action, the common static energy density becomes
\begin{equation}
	\mathcal E_{\rm stat}
	=
	\frac{1}{2}\theta_z^2
	+
	\frac{1}{2}\sin^2\theta\,\varphi_z^2
	+
	D\sin^2\theta\,\varphi_z
	+
	m^2\sin^2\theta
	+
	h\cos\theta .
\end{equation}
Completing the square gives
\begin{equation}
	\mathcal E_{\rm stat}
	=
	\frac{1}{2}\theta_z^2
	+
	\frac{1}{2}\sin^2\theta(\varphi_z+D)^2
	+
	\frac{\delta}{2}\sin^2\theta
	+
	h\cos\theta,
\end{equation}
This is the effective static functional used in the main text. The antiferromagnetic and ferromagnetic theories differ only in how this common static energy is completed dynamically.

\section{Extended pole-crossing profiles}
\label{app:pole-crossing}

The main text focuses on ordinary polar Q-balls, whose coordinate
\(\theta(z)\) remains inside \(0\leq\theta\leq\pi\). The reduced equations also
admit finite-energy profiles in which \(\theta\) is continued beyond this
interval. These configurations start and end at the same physical pole, but
cross the opposite pole. We call them extended pole-crossing profiles.

For a north-pole subtraction, the effective potential can be written in the
unified form
\begin{equation}
V_N(\theta)
=
\frac{A}{2}\sin^2\theta+B(\cos\theta-1)
=
2\sin^2\frac{\theta}{2}
\left(
A\cos^2\frac{\theta}{2}-B
\right),
\label{eq:app-pole-crossing-potential}
\end{equation}
where
\begin{equation}
(A,B)=(\delta,h),\qquad
(A,B)=(\alpha,h),\qquad
(A,B)=(\delta,\Lambda)
\end{equation}
for the static, antiferromagnetic rotating, and ferromagnetic rotating
problems, respectively. A north-pole extended profile satisfies
\[
\lim_{z\to-\infty}\theta(z)=0,
\qquad
\lim_{z\to+\infty}\theta(z)=2\pi .
\]
Since the trajectory must cross the opposite pole,
\[
V_N(\pi)=-2B
\]
must be positive. Thus \(B<0\). Writing \(b=-B>0\), the remaining endpoint
condition is \(A+b>0\), and the profile is
\begin{equation}
\theta_{\rm ext}^{N}(z)
=
\pi+
2\arctan
\left[
\sqrt{\frac{b}{A+b}}\,
\sinh\!\left(\sqrt{A+b}\,(z-z_0)\right)
\right],
\label{eq:app-pole-crossing-profile}
\end{equation}
with the continuous branch chosen so that the solution runs from \(0\) to
\(2\pi\). The south-pole extended profile is obtained by reflection and exists
for
\[
B>0,
\qquad
A+B>0 .
\]

These conditions place the extended profiles in the complementary sign sector
to the ordinary polar Q-balls. Indeed, when the ordinary north-pole Q-ball
exists its condition is \(0<B<A\), whereas the north-pole extended profile
requires \(B<0\). Similarly, the ordinary south-pole Q-ball exists for
\(-A<B<0\), whereas the south-pole extended profile requires \(B>0\). This
complementarity concerns the sign of the effective tilt \(B\); the endpoint
conditions \(A\pm B>0\) still have to be imposed.

The observables can be expressed through two elementary integrals,
\begin{equation}
\mathcal I_{\Omega}(A,b)
=
\int_{-\infty}^{\infty}
(1-\cos\theta_{\rm ext})\,dz,
\qquad
\mathcal I_s(A,b)
=
\int_{-\infty}^{\infty}
\sin^2\theta_{\rm ext}\,dz .
\end{equation}
For \(A>0\),
\begin{equation}
\mathcal I_{\Omega}(A,b)
=
\frac{4}{\sqrt A}\,
\operatorname{arsinh}\sqrt{\frac{A}{b}},
\qquad
\mathcal I_s(A,b)
=
4\left[
\frac{\sqrt{A+b}}{A}
-
\frac{b}{A^{3/2}}
\operatorname{arsinh}\sqrt{\frac{A}{b}}
\right].
\end{equation}
The continuous limit gives the case \(A=0\). For \(-b<A<0\), the same
expressions are understood by analytic continuation, with the hyperbolic
inverse sine replaced by the corresponding inverse sine.

The static branches correspond to \(A=\delta\) and \(b=|h|\). In the polar
regime \(\delta>0\), the north-pole extended profile exists for \(h<0\), while
the south-pole extended profile exists for \(h>0\). Their physical energy is
\begin{equation}
E_{\rm ext}^{\rm stat}
=
\delta\,\mathcal I_s(\delta,b)
+
2b\,\mathcal I_{\Omega}(\delta,b).
\end{equation}

For antiferromagnetic dynamics,
\[
A=\alpha=\delta-\omega^2,
\qquad
b=|h|.
\]
The north-pole branch exists for \(h<0\), \(\alpha-h>0\), while the south-pole
branch exists for \(h>0\), \(\alpha+h>0\). The charge and physical energy are
\begin{equation}
Q_{\rm ext}^{\rm AF}
=
\omega\,\mathcal I_s(\alpha,b),
\qquad
E_{\rm ext}^{\rm AF}
=
\delta\,\mathcal I_s(\alpha,b)
+
2b\,\mathcal I_{\Omega}(\alpha,b).
\end{equation}

For ferromagnetic dynamics,
\[
A=\delta,
\qquad
b=|\Lambda|.
\]
In the polar regime \(\delta>0\), the north-pole extended profile exists for
\(\Lambda<0\), while the south-pole extended profile exists for
\(\Lambda>0\). The corresponding Berry charges and physical energies are
\begin{equation}
Q_{\rm ext,N}^{\rm FM}
=
\mu\,\mathcal I_{\Omega}(\delta,|\Lambda|),
\qquad
E_{\rm ext,N}^{\rm FM}
=
4\sqrt{\delta+|\Lambda|}
-
\frac{h}{\mu}Q_{\rm ext,N}^{\rm FM},
\end{equation}
and
\begin{equation}
Q_{\rm ext,S}^{\rm FM}
=
\mu\,\mathcal I_{\Omega}(\delta,|\Lambda|),
\qquad
E_{\rm ext,S}^{\rm FM}
=
4\sqrt{\delta+|\Lambda|}
+
\frac{h}{\mu}Q_{\rm ext,S}^{\rm FM}.
\end{equation}

At fixed ferromagnetic Berry charge these extended profiles are energetically
above the ordinary polar Q-balls:
\begin{equation}
E_{\rm ext}^{\rm FM}(Q)-E_{\rm ord}^{\rm FM}(Q)
=
4\sqrt{\delta}
\left[
\coth\left(\frac{\sqrt{\delta}\,Q}{4\mu}\right)
-
\tanh\left(\frac{\sqrt{\delta}\,Q}{4\mu}\right)
\right]
>0 .
\end{equation}
{ In particular, a full spectral and dynamical stability analysis beyond the reduced one-dimensional helical ansatz would help determine which of these exact stationary branches can persist as long-lived magnetic textures.
}

\bibliographystyle{apsrev4-1}
\bibliography{references}

\end{document}